%% file: ms.tex
\documentclass[onecolumn, amsmath,amssymb, aps,12pt]{revtex4}
\usepackage{setspace}
\usepackage[dvipsnames]{xcolor}
\usepackage{graphicx}% Include figure files
\usepackage{dcolumn}% Align table columns on decimal point
\usepackage{bm}% bold math
\DeclareUnicodeCharacter{0300}{{\`u}}
\begin{document}

\title{Nonreciprocal guided waves in presence of swift electron beams }% Force line breaks with \\

\author{Asma Fallah\textsuperscript{1}}

\author{Yasaman Kiasat\textsuperscript{1}}
 \email{yasaman.kiassat@gmail.com; address: 210 Locust St., 16E, Philadelphia, Pennsylvania 19106, USA.}
 \author{M{\'a}rio G. Silveirinha\textsuperscript{2}}
 \author{Nader Engheta\textsuperscript{1}}%
 \email{engheta@seas.upenn.edu}
\affiliation{%
 \textsuperscript{1} University of Pennsylvania, Department of Electrical and Systems Engineering, Philadelphia, Pennsylvania 19104, USA
}%
\affiliation{%
\textsuperscript{2} University of Lisbon–Instituto Superior Técnico and Instituto de Telecomunicações, Avenida Rovisco Pais, 1, 1049-001 Lisboa, Portugal}%

\newpage
\begin{abstract}
Breaking the reciprocity of electromagnetic interactions is of paramount importance in photonic and microwave technologies, as it enables unidirectional power flows and other unique electromagnetic phenomena. Here we explore a method to break the reciprocity of electromagnetic guided waves utilizing an electron beam with a constant velocity.  By introducing an effective dynamic conductivity for the beam, we theoretically demonstrate how nonreciprocal guided waves and a one-way propagating regime can be achieved through the interaction of swift electrons with electromagnetic waves in two-dimensional (2D) parallel-plate and three-dimensional (3D) circular-cylindrical waveguides. Unlike the conventional electron beam structures such as traveling wave tubes and electron accelerators, here the goal is neither to generate and/or amplify the wave nor to accelerate electrons.  Instead, we study the salient features of nonreciprocity and unidirectionality of guided waves in such structures.  The relevant electromagnetic properties such as the modal dispersion, the field distributions, the operating frequency range, and the nonreciprocity strength and its dependence on the electron velocity and number density are presented and discussed. Moreover, we compare the dispersion characteristics of waves in such structures with some electric-current-based scenarios in materials reported earlier. This broadband tunable magnet-free method offers a unique opportunity to have a switchable strong nonreciprocal response in optoelectronics, nanophotonics, and $THz$ systems. 
\end{abstract}

\maketitle
 
The reciprocity theorem for electromagnetic fields and waves is a generalization of Lord Rayleigh's reciprocity theorem for sound waves \cite{rf1}. The electromagnetic version of the theorem stems from the seminal work of Lorentz and Helmholtz, and is a cornerstone of the Maxwell theory. The Lorentz reciprocity theorem states that in linear time-invariant structures with materials described by symmetric tensors, one can interchange the locations of the source and observer without changing the observed field strength \cite{rf2,rf3,rf4,rf5}. Achieving nonreciprocity has been the subject of research interest for years, offering practical opportunities over the past decades (e.g., \cite{rf18,rf29,rf8,rf6,rf30,rf27,rf25,rf19,rf13,rf11,rf15,rf21,rf26,rf24,davoyan2014electrically,rf7,rf10,rf14,rf16,rf12,rf17,rf5,bliokh2018electric,rf28,rf23,rf22,rf20,rf9}). For example, a nonreciprocal response is indispensable for realizing circulators, isolators \cite{rf5,rf6,rf7,rf8,rf9,rf10,rf11,rf12,rf13,rf14}, optical diodes \cite{rf15,rf16,rf17,rf18,rf19}, and energy sinks \cite{rf20,rf21}. Therefore, there has been a tremendous effort in the development of robust solutions to break reciprocity \cite{rf4,rf5,rf6,rf7,rf8,rf9,rf10}. The known approaches can, generally, be divided into two categories; (1) solutions that involve DC biasing magnetic field, and (2) the magnet-free solutions.

Reciprocity can be broken via magnetically biased gyromagnetic materials (known since 1950's \cite{rf31,rf32,rf33,rf34}). In addition to its bulkiness, a primary problem with this method is a fairly weak nonreciprocal response at terahertz and higher frequencies, as a Tesla-level magnetic field results in the cyclotron frequency in the microwave frequency range. Indeed, it is usually impractical to push the cyclotron frequency to the $THz$ range or higher frequencies. Furthermore, large magnetic fields require bulky solenoids, which are challenging to integrate with planar technologies and nanophotonics \cite{rf35}.

Several approaches have been developed to break reciprocity without a biasing magnetic field. For instance, nonlinear effects \cite{rf15,rf16,rf17,rf18,rf19,rf29,rf30}, optomechanical interaction \cite{rf36,rf37,rf38}, spatiotemporally modulated guided-wave structures \cite{rf6,rf7,rf9,rf10,rf11,rf12,rf14,rf24,rf24a}, transistor-based metamaterials \cite{rf25,rf26,rf27}, and moving media \cite{rf28,rf36}. These approaches have their own advantages and constraints (e.g., see \cite{rf39,rf39a,rf39b}).

Recently, several methods explored the possibility of using a drift current in graphene to break the time-reversal symmetry \cite{rf20,rf40,rf42g,rf41,rf42}. The drift current originates a "plasmonic" drag effect that can lead to unidirectional propagation regimes \cite{rf40,rf42} and a strong nonreciprocity. The effect of the drift current on the graphene conductivity has been modeled using linear response (Kubo's) theory either through a nonequilibrium Fermi distribution or through a suitable interaction Hamiltonian \cite{rf41,rf42,rf42a}. The effect of drift current has also been studied in semiconductors for a purpose of parametric amplification \cite{rf44}, however, comparing with graphene the velocity of the electrons is constrained by the lower mobility in semiconductors. Moreover, the nonreciprocity due to electric currents in metals has also been investigated \cite{bliokh2018electric}, but here the velocity of electrons is extremely low, resulting in weak nonreciprocity. The nonreciprocal response stems from the fact that the electric current is odd under a time-reversal. Although graphene may exhibit high electron mobility, even in this case the drift velocity is fundamentally limited by graphene's Fermi velocity $v_F=c/300$ \cite{rf43}. The nonreciprocity strength is determined by the Doppler shift $v_0 k$ ($k$ is the plasmons wavenumber, parallel or antiparallel to the direction of motion of electrons, and $v_0$ is the drift velocity). A strong nonreciprocity requires that the Doppler shift $v_0 k$ must be a significant fraction of the operating frequency. Therefore, the nonreciprocal response in graphene is constrained by its Fermi velocity (and in semiconductors and metals such nonreciprocity is even much weaker). Moreover, large drift velocities are very challenging to achieve and can cause heating and compromise the integrity of the graphene sheet, solid-state, and metallic materials.

In this work, we explore a different tunable method to break the reciprocity. Rather than considering that the electrons move in a solid-state material, as in graphene, or in metal, here we suppose that the swift electrons are accelerated in a vacuum to constant high velocities. This approach tackles the aforementioned challenges of achieving high electron velocities. For example, in the cathodoluminescence microscopes, a $30$-$keV$ electron beam has an electron velocity of around  $v_0=c/3$ \cite{CLparam,polman2019electron}. Furthermore, as we show here, the interaction of a beam of swift electrons with guided electromagnetic waves around the electron beam can lead to extreme nonreciprocity at high frequencies and relatively wide bandwidths.  It is worth noting that although the interaction of electromagnetic waves with electron beams have been studied for decades in microwave generators such as the traveling wave tubes \cite{TWT} and in electron accelerator structures \cite{eaccel}, the goal here is neither to generate or amplify waves nor to accelerate electrons. Instead, we investigate quantitatively how tunable broadband nonreciprocity can be achieved in guided waves in presence of electron beams with constant velocities. It should also be noted that the current state-of-art technology may enable the integration on a chip of the circuitry required to generate the electron beam \cite{rf43a,sapra2020chip}. 

We study the interactions between the swift electrons and the transverse-magnetic $TM$ electromagnetic mode in two illustrative cases; (1) a parallel-plate waveguide containing an electron sheet (planar beam) with electron velocity $v_0$, to demonstrate the theory for the two-dimensional (2D) scenario; and (2) a hollow metallic cylindrical uniform waveguide with a circular cross section having a collimated electron beam (pencil beam) along its axis, as a possible case for future experimentation. We theoretically study the dispersion relations for these two cases and examine the nonreciprocal response of the corresponding waveguides for various electron beam constant velocities. Our proposal can obviously be extended to other waveguide geometries. In the following, a time-harmonic convention of the form $e^{i\omega t}$, with $\omega$ as the operating angular frequency, is assumed.

%\section{Theory}
%\subsection{Modeling of Electron Beam}

\begin{figure}
    \centering
      \includegraphics[width=10cm]{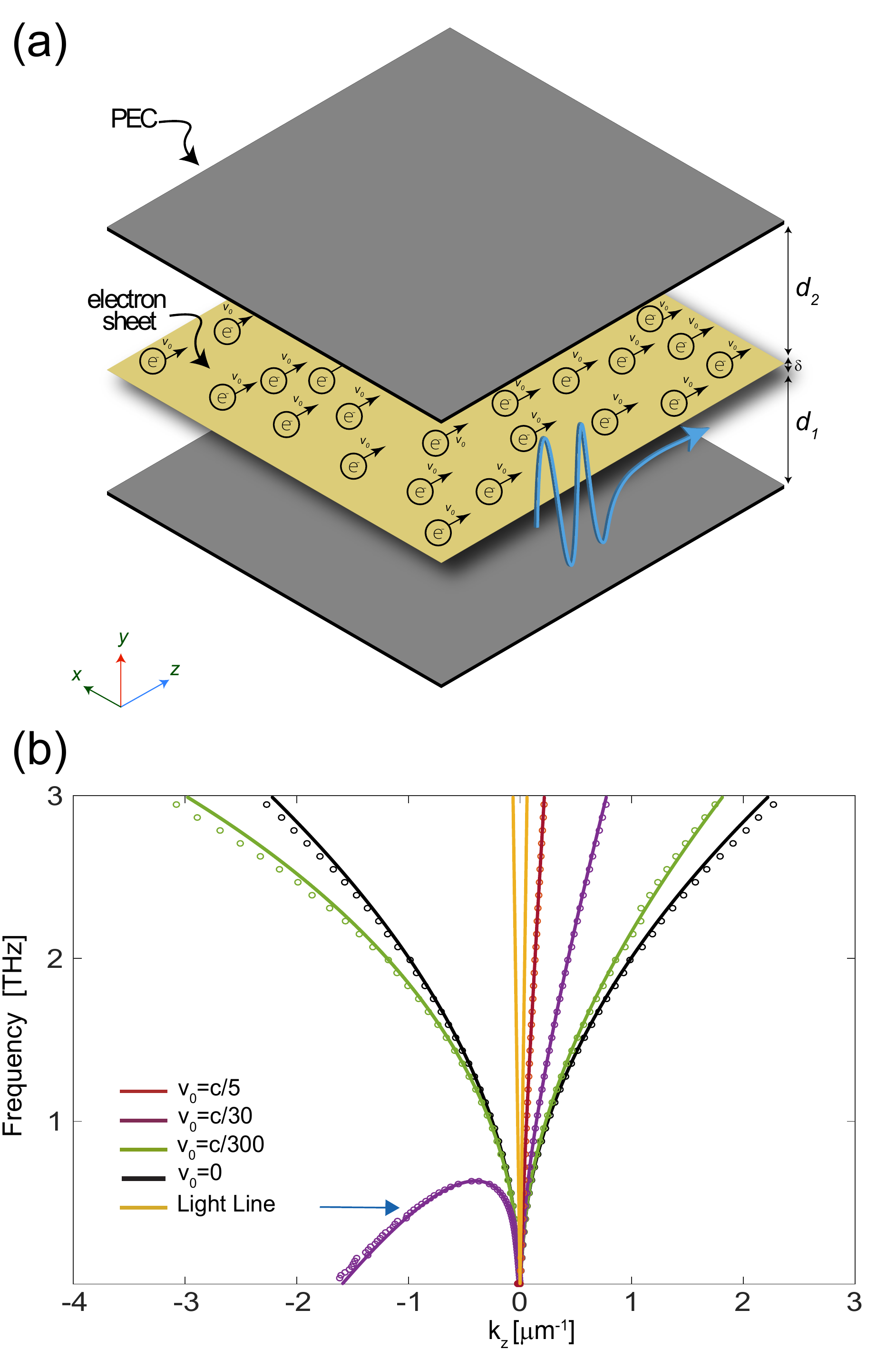}
    \caption{(a) Schematic of the parallel-plate two-dimensional (2D) waveguide containing a two-dimensional (planar) electron beam (yellow region). (b) Dispersion diagram for the $TM_{10}$ mode of the parallel plate waveguide for various electron velocities $v_0$. Solid lines: analytical results; circles: simulation results. The blue arrow indicates the frequency $f=0.48$ $THz$ corresponding to the electric field profiles shown in the supplementary Fig. S1 \cite{sm}.}
    \label{fig:struc}
\end{figure}

It is worth noting here that the electron beam is assumed to be associated with a steady flow (stationary current) of electrons in vacuum inside the guided-wave structure with cross section invariant along the axis, and consequently, it does not lead to the emission of Cherenkov type radiation \cite{rf43b} or Smith-Purcell type radiation \cite{smith1953visible}, and creates only a static-type field distribution.

The current density of the electron beam can, however, be perturbed by a high-frequency (external) electromagnetic field propagating in the guide. The perturbation can be characterized using a transport equation obtained with the Boltzmann’s formalism (see Eq. S1  \cite{sm}). It is worth noting that the Boltzmann equation, which is usually utilized for charged carriers in materials, has been applied here to the stream of electrons in vacuum.  (In the supplementary materials, we have also included another derivation that regards the electron beam as a polarizable rod, rather than as a bulk material; the two approaches lead to the same physics.) It is possible to linearize the continuity (see Eq. S2 \cite{sm}) and the transport equations. The electron density and the electron velocity are split into the equilibrium (DC) part and time-harmonically varying (AC) part due to the perturbation by an external field. The AC terms are treated as small perturbations of the corresponding DC values (e.g., products of the AC parts of two quantities are neglected). In this manner, it is possible to show that the longitudinal AC current density (Eq. S3 and Eq. S6 \cite{sm}) is linked to the AC field as follows:

\begin{equation}
J_{z}=\frac{\left(q^{2} n_{0} / m\right) \omega E_{z}}{i\left[\left(\omega-v_{0} k_{z}\right)^{2}-k_{z}^{2} \beta\right]+\tau^{-1} \omega},
\label{eq:jtot}
\end{equation}

where the electron beam is assumed to travel along the $z$ axis (Fig. \ref{fig:struc}(a)), $E_{z}$ and $k_z$ are, respectively, the longitudinal $z$-components of the time-varying electric field at the electron beam and wave vector of the $TM$ mode of the waveguide, and $n_0$, $m$, $q$, $v_0$ are all constant quantities, denoting the equilibrium electron number density, electron rest mass, the electron charge, and constant electron velocity, respectively. Furthermore, $\tau$ and $\beta$ represent the average scattering time and diffusion coefficient (e.g., due to electron-electron repulsion), respectively. We point out that Eq. \ref{eq:jtot} describes the effective dynamic conductivity of the electron beam.  In other words, it gives the effective medium (macroscopic) conductivity for the time-varying monochromatic parts of the electric field and current density. Since the electron beam thickness is assumed to be very small, the transverse components of the electric field (if any) do not interact with the beam. Thus, only the longitudinal component of electric field interacts noticeably with the electrons. It is worth noting that for velocities very close to the speed of light $c$, a relativistic correction (Lorentz factor) for the electron mass should be considered \cite{jackson1999classical}. However, in our results reported in this Letter, the maximum velocity we consider is $c/3$, for which the Lorentz factor, being 1.0607, is still very close to unity.  To highlight the main features of the guided modes, in the following we neglect the diffusion and scattering terms. This leads to,

\begin{gather}
    J_z=\ \frac{ q^2 n_0 \omega E_{z}}{{im}(\omega - v_0k_z)^2 }=-i \omega\epsilon_0 (\frac{\omega _p }{ \tilde{\omega} })^2 E_{z},
    \label{eq:j}
\end{gather}

where $\omega_p^2 \equiv q^2 n_0/m \epsilon_0 $ and $\tilde{\omega}\equiv(\omega - v_0k_z)$ which is the Doppler-shifted frequency in the frame co-moving with the electrons. Equation \ref{eq:j} reveals that from the point of view of the dynamic (i.e., AC part) electromagnetic signal, the effective permittivity of the beam (in the laboratory frame) is alike to a Drude model, with the frequency term replaced by $\tilde{\omega}$. In particular, the model predicts that the beam response is different for oppositely-signed values of $k_z$, i.e., for the two opposite directions of guided-wave propagation for a fixed operating frequency. This property is a clear fingerprint of break of reciprocity.

\begin{figure}
    \centering
      \includegraphics[width=10cm]{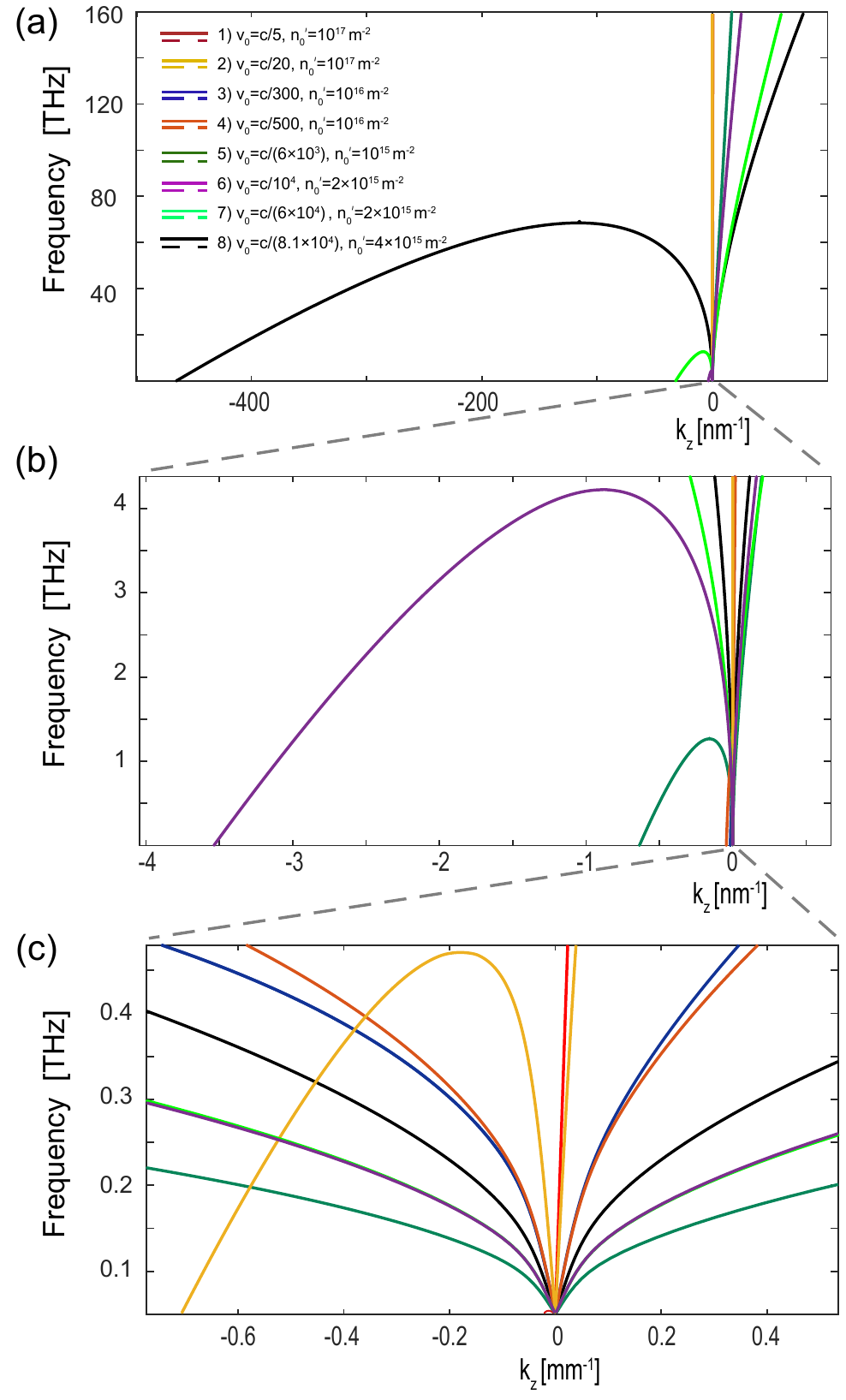}
    \caption{ Comparative study:  Dispersion diagrams of the parallel plate waveguides with thickness of $d=0.05$ $mm$ and various values for the electron beam properties ($n_0'$ and $v_0$). Three of the eight sets of parameters are for specific materials and five others are various parameters arbitrarily chosen. Cases (3), (5), and (8) are dispersion diagram of this waveguide with the graphene \cite{rf40}, GaAs \cite{rf44}, and InSb \cite{sydoruk2010terahertz}, respectively. From panel (a) towards panel (c), zoomed-in plots near the origin are shown.}
    \label{fig:pspp}
\end{figure}

%\subsection{Interaction of Electron Beam with Electromagnetic Field}

To highlight the consequences of the nonreciprocal interactions between the beam and the wave, first, we consider a (2D) parallel-plate waveguide, which, without loss of generality, is assumed to be made of perfect electric conducting (PEC) plates. (For this 2D scenario the electron beam is in the $xoz$ plane (planar beam), and thickness of the beam along $y$ is assumed to be negligible (thin planar beam)). The electron beam travels in a direction parallel to the PEC plates, as shown in Fig. \ref{fig:struc}(a). The distance between the metallic plates is $d$, while the distance between the planar beam and the bottom and top PEC plates is $d_1$ and $d_2$, respectively. Solving Maxwell’s equations and applying the appropriate boundary conditions at the waveguide's plates and at the electron sheet, the modal dispersion and the corresponding electromagnetic field distributions can be analytically derived (details are shown in the supplementary materials). The electron beam is surrounded by a vacuum and its thickness, $\delta$, is considered to be electrically thin enough so $E_{z}$ is assumed to be approximately uniform across the beam.

For the $TM_{10}$ mode with $E_z$, $E_y$, and $H_x$ field components and denoting $k_0^2 \equiv \omega^2 \mu_0 \epsilon_0$ and $ k_y^2 \equiv k_0^2-k_z^2,$ the dispersion relation is given by,

%\begin{gather}
 %   (\frac{\omega_p}{\tilde{\omega}})^2=\frac{-2}{k_y \tan{\frac{k_y}{2}d}} \label{eq:drppw},
  %  \end{gather}
\begin{equation}\frac{q^2 n_0'/m \epsilon_0}{\tilde{\omega}^2}=-\frac{1}{k_{y} \tan (k_y d_1)}-\frac{1}{k_{y} \tan (k_y d_2)}.\label{eq:drppw} \end{equation}

Note that since $\delta$ is infinitesimally small, here for the planar case we define a surface density for the electron beam to be $n_0'=n_0\delta$ which is expressed as an electron number density per unit area.
 
Figure \ref{fig:struc}(b) shows the calculated dispersion diagrams, for $d_1=d_2=d/2$ with $d=0.05$ $mm$, and $n_0'=10^{17}$ $ m^{-2}$. The solid lines represent the dispersion curves obtained from our analytical expression (Eq. \ref{eq:drppw}), while the circles show the results of numerical simulations done using the commercially available COMSOL Multiphysics\textsuperscript{\textregistered} \cite{rf45} (see the supplementary materials for the simulation methods \cite{sm}).  The simulation and analytical results show that the higher the electron velocity is, the larger is the spectral asymmetry and the difference
between the wave-numbers of the guided waves (the $TM_{10}$ mode) propagating in $+z$ and $-z$ directions. Thus, the simulation and analytical results show unequivocally that the motion of electrons in the electron beam inside the guide can result in Fresnel-type drag of the electromagnetic waves, and lead to a very strong nonreciprocity even at high frequencies.  In order to highlight the strength of nonreciprocity in our high-velocity electron beam structures, in figure \ref{fig:pspp} as a parametric study we present dispersion diagrams for several values of electron number densities and electron velocities, some of which are the parameters for the cases of GaAs ($n_0'=10^{15}$ $m^{-2}$, $v_0=5 \times 10^4$ $m/s$ \cite{rf44}), InSb ($n_0'=4 \times 10^{15}$ $m^{-2}$ , $v_0=3.7 \times 10^3$ $m/s$ \cite{sydoruk2010terahertz}), and graphene  ($n_0'=10^{16}$ $m^{-2}$, $v_0=v_F$ \cite{rf40}) reported in the literature.  In this figure, panels (b) and (c) show the zoom-in versions of panels (a) and (b), respectively, around the origin.  It is important to note that since in our structure the electron velocity can be much higher than those in GaAs, InSb, and graphene (as here the electron velocity is not constrained by the mobility of electrons in those materials), much stronger nonreciprocity and unidirectionality can be obtained.  Comparison for the case of cylindrical waveguides is discussed in the following section.  As shown in Fig. \ref{fig:struc}(b) and Fig. \ref{fig:pspp}, at some frequencies the dispersion diagrams bend in opposite directions (up or down) depending on the sign of $k_z$. The turning points depend on the electron velocity among other parameters. Clearly, as can be seen in Fig. \ref{fig:pspp} the strength of nonreciprocity can be tailored with the electron velocity and the electron density. This is a rather unique and attractive feature of this magnetic-free nonreciprocal platform where the electron flow with constant velocity breaks the time reversal symmetry.

 In addition, in figure S1 \cite{sm} of supplementary material we have depicted time snapshots of the $y$- and $z$-components of the electric field distributions at $f=0.48$ $THz$ when $v_0=c/30$.   It is worth mentioning that at this frequency a hollow waveguide with the same thickness would be well below the cut-off frequency for the $TM_{10}$ mode if we did not have the moving electron beam.  The plasma type response of the electron beam suppresses the cut-off of this mode, even in the limit $v_0=0^+$). As it is seen in  Fig. \ref{fig:struc}(a), at the selected frequency of operation, there are three allowed values for the longitudinal wavenumbers; one positive $k_z$ and two negative $k_z$ values. In Fig. S1, we present the modal field distributions for the three possible $k_z$ values at $0.48$ $THz$. Note that one of the negative $k_z$ modes is a backward wave. From Fig. S1, it is evident that the field profiles are rather different for the $+z$ and $-z$ directions of propagation, which is another manifestation of the nonreciprocity of the system. This result can also be intuitively explained by the fact that according to Eq. \ref{eq:j} the effective dynamic conductivity of the electron beam explicitly depends on $k_z$ when $v_0 \neq 0$.       

\begin{figure}
  \centering
 \includegraphics[width=10cm]{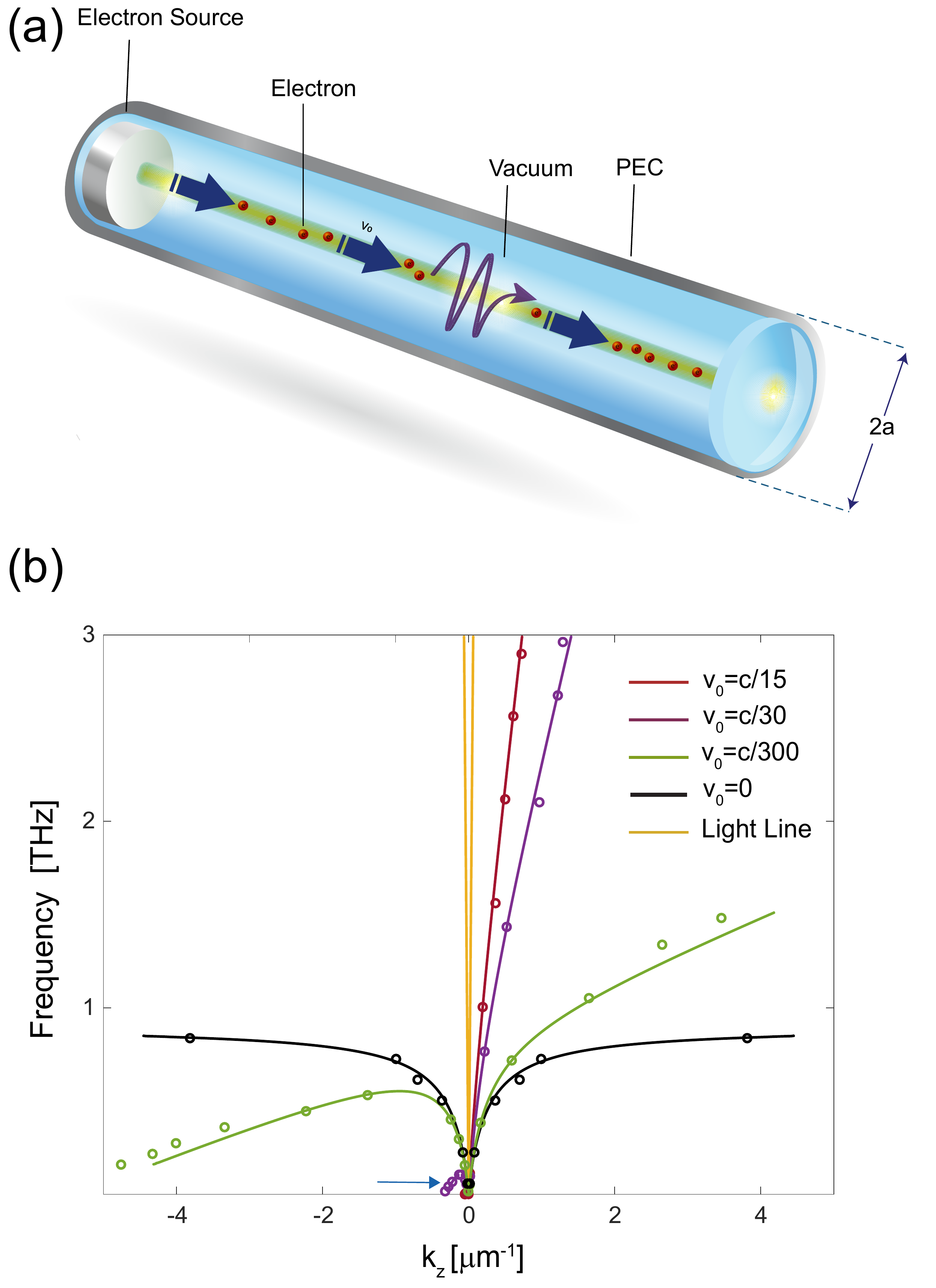} 
\caption{(a) Schematic of the circular cylindrical waveguide with an axial electron pencil beam; the beam trajectory is represented by the green line. (b) Dispersion diagram for the $TM_{10}$ mode of cylindrical waveguide for various beam velocities $v_0$. Solid lines: analytical results; circles: simulation results. The blue arrow marks the frequency $f=0.08$ $THz$ corresponding to the electric field profiles shown in Fig. S2 \cite{sm}. }
\label{fig:geomdrcyl}
\end{figure}
\begin{figure}
\includegraphics[width=10cm]{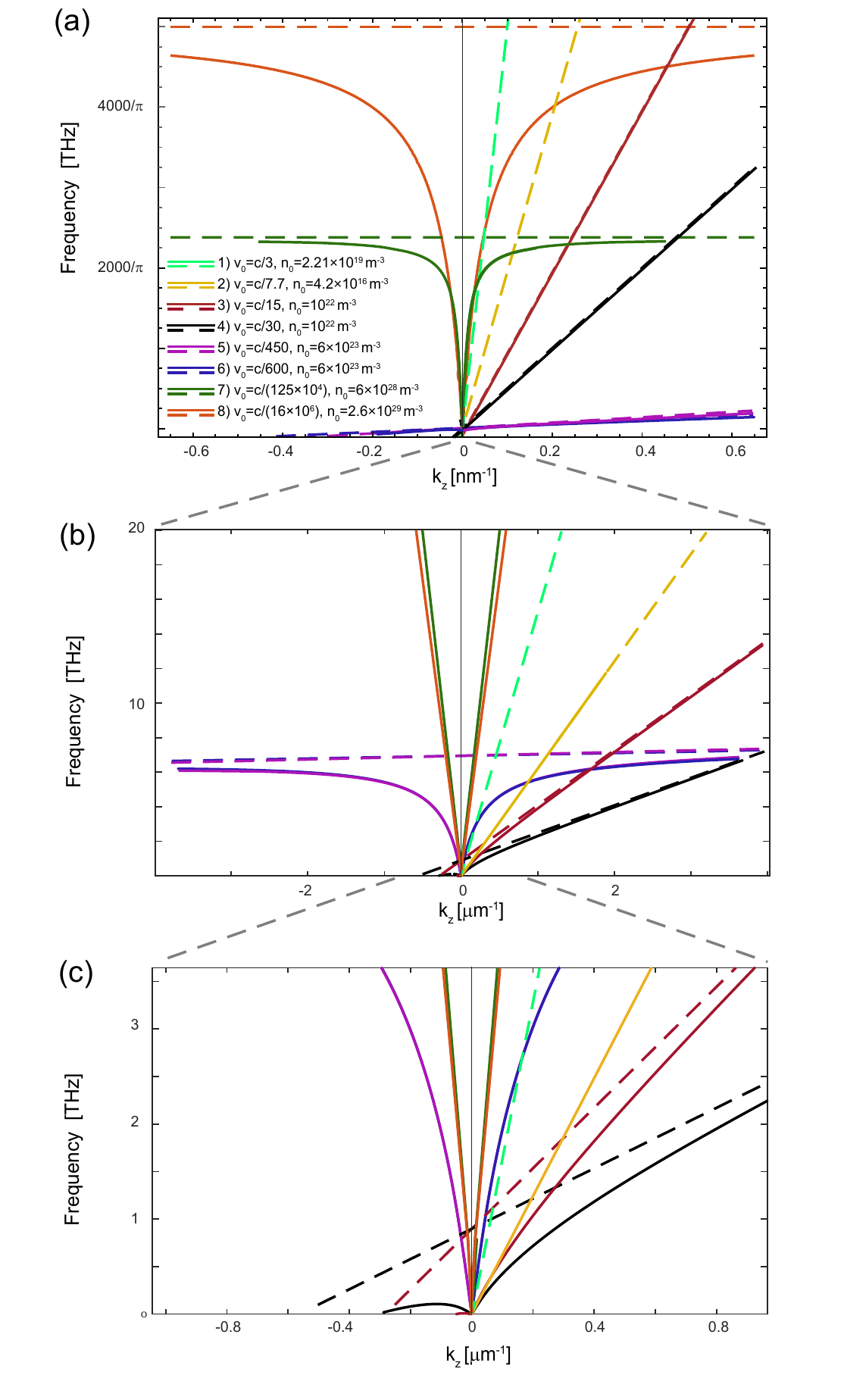}
\caption{\label{fig:pscyl}Comparative study:  Dispersion diagrams of the cylindrical waveguides with an outer radius of $a=40$ $\mu m$
and various values for the electron beam properties ($n_0,v_0$ and $b$). Four of the eight sets of parameters are for specific cases of materials and electron beams and four others are parameters arbitrarily selected. Cases (1), (2), (7), and (8) are dispersion diagram of the cylindrical waveguide containing the cathodoluminescence beam \cite{CLparam,polman2019electron}, an electron gun (EM 503) \cite{egun}, gold rod \cite{davoyan2014electrically}, and a metal \cite{bliokh2018electric}, respectively. Solid and dashed lines represent the dispersion diagrams and their asymptotes, respectively.  The radii of the electron beams for cases (1) and (2) are $3$ $nm$ and $1.5$ $\mu m$, for (3-6) $3.8$ $\mu m$, and for (7) and (8) $0.1$ $\mu m$ and $20$ $nm$, respectively. From panel (a) towards panel (c), zoomed-in plots near the origin are shown.}
\end{figure}

Next, we analyze the case of the 3D hollow circular-cylindrical waveguide of radius $a$ with PEC wall, with a pencil-type electron beam of radius $b$ flowing with constant velocity $v_0$ along this waveguide axis (Fig. \ref{fig:geomdrcyl}(a)). The dynamic (i.e., AC part) current density in the electron beam is still given by Eq. \ref{eq:j}, but now we use $n_0$ which is number density per unit volume. To investigate the nonreciprocal behavior of this system, we analytically derived the dispersion relation for the $TM_{10}$ mode. Having $ k_r^2 \equiv k_0^2-k_z^2,$  the dispersion relation of the $TM_{10}$ mode supported by the system in the limit of $b\ll \lambda$ is determined by,

\begin{gather}
    (\frac{\omega_p}{\tilde{\omega}})^2 \approx - \frac{2}{k_rb} \frac{C(k_r a)J_1\; (k_rb)+Y_1\; (k_rb)}{C(k_ra)J_0(k_rb)+Y_0(k_rb)}+1,
    \label{eq:dr}
    \end{gather}
    where $C(k_ra) \equiv - Y_0(k_ra)/J_0(k_ra)$ with $J_n$ and $Y_n$ being the Bessel functions of the first and second kind and $a$ is radius of the cylindrical waveguide.

Figure \ref{fig:geomdrcyl}(b) depicts the dispersion diagram (solid lines: analytical results, circles: numerical results) of the $TM_{10}$ mode for several electron velocities $v_0$ ($v_0=c/15, c/30, c/300$) and considering $n_0=10^{22}$ $m^{-3}$, $a=38$ $\mu m$, and $b=3.8$ $\mu m$. As before, it is found that the electron flow in the channel results in large spectral asymmetry and strong nonreciprocity. For instance, for $v_0=c/300$ and for frequencies above $0.4$ $THz$ the wave can experience highly nonreciprocal behavior.
Similar to the parallel plate case, there is also a "turning-point frequency" where the group velocity vanishes. For frequencies below the turning point, the guide supports modes with a negative $k_z$ value. It is worth noting that for the electron beam parameters of Fig. \ref{fig:geomdrcyl}, the total current is much larger than the currents than can be provided by some of the conventional electron beam devices such as the cathodoluminescence microscope. The reason we use a large current is to highlight more clearly the impact of the electron beam on the wave propagation. In a cathodoluminescence microscope the typical currents are in the range of $10$ $pA$ and $10$ $nA$ (e.g. see \cite{CLparam,polman2019electron}) for electrons with velocity $c/3$. Hence, to show the compatibility of our proposal with commercially available electron beam generators along with other practical and theoretical cases, we conducted a parametric study for different scenarios. Fig. \ref{fig:pscyl} demonstrates dispersion diagrams of the cylindrical waveguide case with various values for the electron beam's properties such as radius, velocity, and density. This study includes four specific practical cases such as metallic wires \cite{davoyan2014electrically,bliokh2018electric}, cathodoluminesence (CL) beam \cite{CLparam,polman2019electron}, and an electron gun \cite{egun}, along with four other cases with selected arbitrary parameter values. It is important to note that for the cylindrical cases here, the asymptotic line of the dispersion curves is $(\omega - v_0k_z)=\omega_p$, which shows that its intersection with the vertical axis is at the plasma frequency and its slope is the velocity of electrons.  This implies that for higher electron velocity and/or lower plasma frequency we can achieve stronger nonreciprocity and unidirectionality.  In order to highlight this point, we have considered and shown several cases with different $n_0$ (thus different plasma frequencies) and different $v_0$. In Fig. \ref{fig:pscyl}, asymptotic lines have been presented with dashed lines.  Comparison among the various cases (including the four practical cases mentioned above) is informative in this figure.  For example in the metal case \cite{bliokh2018electric} the plasma frequency is very high and the electron velocity is very low, and consequently the electric-current-based nonreciprocity in metals is very weak. As we decrease the plasma frequency and increase the electron velocity, i.e., for cases of interest to us, e.g., for the CL and electron gun beams, the nonreciprocal response and unidirectionality become much stronger over a wide range of frequencies. Interestingly, in these cases, the response can be strongly nonreciprocal such that the turning point of the modal dispersion occurs at very low frequencies, yielding thereby a broadband regime of unidirectional propagation.

It is worth noting that in all these cases, the average longitudinal distance ($d_{avg}$) between neighboring electrons in the beam must be less than the wavelength of guided wave along the beam so that the effective theory remains valid. Here one can consider $d_{avg}=1/n_0A$ where $A=\pi b^2$  is cross-sectional area of the electron beam. In other words, the frequency of operation should stay below the frequency for which the longitudinal wavenumber $k_z$ is less than  $k_{avg}=2 \pi /d_{avg}=2 \pi n_0 A$. For instance, for the electron gun (EM 503) the $k_{avg}$ is $1.86$ $\mu m^{-1}$, therefore frequencies with wavenumber smaller than  $1.86$ $\mu m^{-1}$ are applicable. The average distance of cases (1) through (8) in Fig. 5 (cases (3,4) and (5,6) have same number density and same radius, hence same $k_{avg}$) are $3.92$ $ mm^{-1}$, $1.86$ $\mu m^{-1}$, $2.85$ $ pm^{-1}$, $170$ $ pm^{-1}$, $11844$ $ pm^{-1}$, and $2052$ $ pm^{-1}$, respectively. Hence, as it is demonstrated in Fig. \ref{fig:pscyl}, for the electron gun and CL cases, dispersion curves (solid lines) have been stopped at the frequencies ($11.5$ $THz$ and $62.3$ $GHz$, respectively) beyond which the effective medium theory will not be valid.

Supplementary Figure S2 \cite{sm} shows the time snapshots of the $r$- and $z$-components of the electric field distributions at $f=0.08 THz$ and $v_0=c/30$.  As before, the fields are highly concentrated near the electron beam. We should also note that the nonreciprocal guided mode is somewhat alike to a guided surface plasmon concentrated near the electron beam, and its propagation is little affected by the outer wall of the waveguide. However, the presence of the waveguide wall ensures that only this nonreciprocal $TM_{10}$ mode can propagate and that the higher-order modes are below the cutoff. See the supplementary materials for the dispersion curves for different waveguide radii \cite{sm}.

%\section{Conclusion}

In conclusion, in this Letter we have discussed a mechanism to break the electromagnetic reciprocity at relatively high frequencies that exploits the interaction of electromagnetic guided waves with electron beams with moderately high constant velocity. We have illustrated the concept by calculating the dispersion of the guided modes of two different hollow metallic waveguides containing a moving electron beam surrounded by a vacuum. Our theoretical analysis has shown that the guided modes can be effectively dragged by the swift electrons and that the light-matter interactions in the guide enable unidirectional propagation regimes and a strong nonreciprocity. We have compared the case of high-velocity electron beams with some of the practical scenarios, and have concluded that the electron beams can provide much stronger nonreciprocity and unidirectionality.  The proposed method opens up new perspectives and possibilities in controlling and manipulating the direction of flow of electromagnetic waves in optoelectronics, $THz$ systems, and nanophotonics networks, as the nonreciprocity strength is not constrained by the material properties.  This form of nonreciprocity can be applicable to other general guided-wave structures with hollow cores such as holey fibers. 

%\begin{acknowledgments}
The authors express their thanks to Miguel Camacho of the University of Pennsylvania for valuable discussions.
%\end{acknowledgments}

This work is supported in part by the National Science Foundation (NSF) Emerging Frontiers in Research and Innovations (EFRI) Program grant $\#$1741693.

\nocite{*}

\bibliography{main}% Produces the bibliography via BibTeX.

\include{supp}

\end{document}

%% file: supp.tex
\setcounter{equation}{0}
\setcounter{figure}{0}
\setcounter{table}{0}
\setcounter{page}{1}
\makeatletter
\renewcommand{\theequation}{S\arabic{equation}}
\renewcommand{\thefigure}{S\arabic{figure}}
\renewcommand{\bibnumfmt}[1]{[SR#1]}
\renewcommand{\citenumfont}[1]{SR#1}

\begin{center}
\textbf{\large Supplementary Materials: "Nonreciprocal guided wave in presence of swift electron beams"}\\
\textit{Asma Fallah\textsuperscript{1}, Yasaman Kiasat\textsuperscript{1},M{\'a}rio G. Silveirinha\textsuperscript{2},Nader Engheta\textsuperscript{1}}\\
\textit{\textsuperscript{1}University of Pennsylvania, Department of Electrical and Systems Engineering, Philadelphia, Pennsylvania 19104, USA}\\
\textit{\textsuperscript{2}University of Lisbon–Instituto Superior Técnico and Instituto de Telecomunicações, Avenida Rovisco Pais, 1, 1049-001 Lisboa, Portugal}\\
\textit{[fallah,engheta]@seas.upenn.edu}
\end{center}

\maketitle

\section{Electron Beam Model}

Here we derive the dynamic (i.e., AC part) of electromagnetic response of the moving electron beam as we treat the beam as a medium with effective dynamic conductivity. Our analysis is based on the following transport equation, which follows from Boltzmann's theory \cite{rf46s,boltzmn2s,boltzmn3s}:

\begin{gather}
\begin{split}
\\
    &
   [\frac{\partial }{\partial t}\left(n\left(\boldsymbol{r},t\right)\textbf{v}\left(\boldsymbol{r},t\right)\right)+
    n\left(\boldsymbol{r},t\right)\left(\textbf{v}\left(\boldsymbol{r},t\right).\mathrm{\nabla }\right)\textbf{v}\left(\boldsymbol{r},t\right)+\textbf{v}\left(\boldsymbol{r},t\right)\mathrm{\nabla }\mathrm{.}(n\left(\boldsymbol{r},t\right)\textbf{v}\left(\boldsymbol{r},t\right))]
    \\
    &
-\frac{q}{m}n\left(\boldsymbol{r},t\right)\textbf{E}\left(\boldsymbol{r},t\right)
+{\beta }\mathrm{\nabla }n\left(\boldsymbol{r},t\right)+\frac{n\left(\boldsymbol{r},t\right)\textbf{v}\left(\boldsymbol{r},t\right)}{{\tau }}=0, 
\end{split}
\label{eq:bltz}
\end{gather}

where the first three terms of the equation denote the convective derivative terms. Furthermore, the fourth term models the electric force acting on the electrons due to the external field. The fifth term represents the diffusion of electrons due to electron-electron repulsion and the last term models possible electron collisions with air molecules. In addition, the electron density and velocity must satisfy the continuity equation (hereafter we drop the label $(\boldsymbol{r},t)$ for the sake of conciseness)
 \begin{gather}
     \frac{\partial n}{\partial t}+\mathrm{\nabla }\mathrm{.}\left(n\textbf{v}\right)=0,
     \label{eq:cont}
 \end{gather}

One can linearize the total current density of electrons by splitting the number density $n=n_0+ \tilde{n}$ and the electron velocity $\textbf{v}=\textbf{v}_0+ \tilde{\textbf{v}}$ into constant ("DC") equilibrium parts and ("AC") (nonequilibrium, i.e., dynamic) parts due to the perturbation by an external field. The linearized AC part of the current density is given by
 
 \begin{gather}
 \tilde{J}=q(n_0\tilde{\textbf{v}}+\tilde{n}\textbf{v}_0).
 \label{eq:jlin}
  \end{gather}

where product of the AC parts of the number of density and velocity were neglected due to small amplitude ($\tilde{n}\tilde{\textbf{v}} \ll n_0 \tilde{\textbf{v}} $ and $ \tilde{n} \textbf{v}_0$). By considering Eq. \ref{eq:bltz} and \ref{eq:cont}, supposing that the electron beam moves along the $z$ axis, for a time-harmonic external field such as $E\sim e^{i \omega t-i k_z z}$, it is found that the oscillating (i.e. "AC") parts of the velocity and number of density of electrons after the linearization satisfy: 

\begin{equation}
\tilde{v}=\frac{q\left(\omega-v_{0} k_{z}\right) E_{z} / m}{i\left[\left(\omega-v_{0} k_{z}\right)^{2}-k_{z}^{2} \beta\right]+\tau^{-1} \omega}, \label{eq:linv}
\end{equation}

\begin{equation}\tilde{n}=n_{0} \frac{ k_{z}  \tilde{v}}{\left(\omega - v_{0}  k_{z}\right)}.  \label{eq:linn0}\end{equation}

 Eq. \ref{eq:linv} and Eq. \ref{eq:linn0} are linearized versions of the transport (Eq. \ref{eq:bltz}) and continuity equations (Eq. \ref{eq:cont}), respectively, where $E_{z}$ and $k_z$ are, sequentially, the longitudinal ($z$-component) of the electric field and wave vector calculated at the electron beam, and $n_0$, $m$, $q$, $v_0$ are all constant quantities, representing the equilibrium (DC) electron number density, mass of the electron, the electron charge, and electron velocity, respectively. Furthermore, as already mentioned, $\tau$  and $\beta$ represent average scattering time and diffusion coefficient, respectively. By substituting Eq. \ref{eq:linv} and Eq. \ref{eq:linn0} into Eq. \ref{eq:jlin}, one finds that the AC current density is related to the AC electric field as follows:
 
\begin{equation}
J_{z}=\frac{\left(q^{2} n_{0} / m\right) \omega E_{z}}{i\left[\left(\omega-v_{0} k_{z}\right)^{2}-k_{z}^{2} \beta\right]+\tau^{-1} \omega}.
    \label{eq:totaljap}
\end{equation}

The coefficient that multiplies the tangential electric field in the right-hand side of Eq. \ref{eq:totaljap} represents the effective dynamic conductivity of the electron beam. 

\section{Effective polarizability model of the electron beam}

In this section, we derive an alternative model for the optical response of the electron beam, where the electron beam is regarded as a body with negligible transverse dimensions. In contrast, the model considered in the main text (as also mentioned in Sec. I here) treats the electron beam as a bulk medium. 
The model discussed here can be especially useful if the electron beam is relatively dilute so that the number of electrons in the cross section is small. In such a case, it seems more appropriate to regard the beam as a linear-type object described by a certain polarizability $\alpha$ per unit of length. The polarizability links the electric dipole moment per unit of length of the linear object ($p_z$) with the incident (dynamical, AC) electric field as $p_z=\epsilon_0 \alpha E_z$. The electric dipole moment is due to perturbations of the motion of the moving electrons caused by the incident field. Such perturbations lead to an oscillatory motion of the electrons with respect to the unperturbed trajectory with constant velocity $v_0$. 

To obtain the effective polarizability, first, we consider the co-moving frame (primed coordinates) where the electrons in the beam are (without the external perturbation) motionless. The electrons are assumed to be placed along the $z^{\prime}$-axis. The average distance between electrons is $d$. We ignore electron-electron ($e-e$) repulsive interactions (which is justified in the dilute limit considered here; such interactions can in principle be modeled with a diffusion term). Under these assumptions, the equation of motion for a single electron is $m \frac{d^{2} z^{\prime}}{d t^{2}}=q E_{z}^{\prime}$  , where $m$ is the electron mass,  $q=-e$ is the electron charge, $E_z^{\prime}$  is the longitudinal electric field in the co-moving frame and $z^{\prime}$   determines the electron coordinates. We neglect relativistic corrections of time so that $t^{\prime}=t$, since even the highest velocity we considered in the main text, i.e., $v_0 = c/3$ has the Lorentz factor, $1.0607$, which is still close to unity.

Using now $d / d t=i \omega^{\prime}$  with  $\omega^{\prime}$ the frequency of the excitation field in the co-moving frame, we see that the induced electric dipole moment per unit of length is given by:

\begin{equation}
p^{\prime}=\frac{1}{d} z^{\prime} q=\frac{1}{d} \frac{-e^{2}}{m \omega^{\prime 2}} E^{\prime}. \label{eq:p1}
\end{equation}

Note that the electric dipole moment is always anti-parallel to the electric field. From here, it is clear that the polarizability evaluated in the co-moving frame is:

\begin{equation}
\alpha_{\mathrm{co}}=\frac{p^{\prime}}{\varepsilon_{0} E_{z}^{\prime}}=\frac{1}{d} \frac{-e^{2}}{\varepsilon_{0} m \omega^{\prime 2}}=-\frac{e I_{D C} }{\varepsilon_{0} m v_{0} \omega^{\prime 2}}. \label{eq:p2}
\end{equation}

Here, $I_{D C}=\frac{e v_{0}}{d}>0$  is the electron beam current. 

The polarizability in the laboratory frame can now be found with a Galilean transformation (which is a good approximation to the Lorentz transformation in the non-relativistic regime considered here). A Galilean transformation gives $p \approx p^{\prime}$  and  $E_{z} \approx E_{z}^{\prime}$  (note that  $E_{z}$ is a field component parallel to the beam motion). Taking into account that the frequencies in the co-moving and laboratory frames are linked by a Doppler shift ($\omega^{\prime}=\omega-k_{z} v_{0}$ ), we finally conclude that $\alpha=\frac{p}{\varepsilon_{0} E_{z}} \approx \frac{p^{\prime}}{\varepsilon_{0} E_{z}^{\prime}}=\alpha_{c o}$ , so that the longitudinal polarizability is given by:

\begin{equation}
\alpha(\omega)=-\frac{e I_{D C}}{\varepsilon_{0} m v_{0}\left(\omega-k_{z} v_{0}\right)^{2}}=-\frac{1}{d} \frac{e^{2}}{\varepsilon_{0} m\left(\omega-k_{z } v_{0}\right)^{2}}.\label{eq:p3}
\end{equation}

This result is consistent with the result of the main text in the collision-less limit. In fact, using Eq. (1) of the main text, one can find that the electric dipole moment per unit of length is given by $p=\int \frac{J_{z}}{i \omega} d s \approx A \times \frac{J_{z}}{i \omega}=\alpha E_{z}$, with $A$ the cross-section of the electron beam and $\alpha$ is defined exactly in the same way as in Eq. \ref{eq:p3} [note that  $A n_{0}=1 / d$].

To conclude, we note that $d$  sets a bound on the spectral region where the “homogenization” applies. One can estimate that the effective theory is valid up to the frequency $f_{\max } \sim \frac{c}{d}=\frac{c}{v_{0}} \frac{I_{D C}}{e}$ .

\section{Derivation of Dispersion relations and Field Distributions}

Here, we derive Eqs. 3 and 4 of the main text that determine the dispersion of the parallel plate and cylindrical waveguide in presence of the electron beam.  In addition, here we also explicitly show the field distributions. In both waveguides shown in Fig. 1(a) and Fig. 3(a),  transverse magnetic ($TM_{10}$) mode has been exploited by solving the Helmholtz equation $ \nabla^2E_{z}+k^2 E_{z}=0$. In the parallel plate waveguide, the Helmhotlz equation can be solved in two regions (i=1 and 2 for regions below and above the electron beam, respectively) by considering $k^2=k_z^2+k_y^2$. Here, $E_{z i}=e_{z i} e^{-i k_{z} z}$ where $e_{z i}$ assumed to be sinusoidal as shown in Eq. \ref{eq:ezpar}.

\begin{equation}
e_{z i}= A_{i} \sin \left(k_{y} y\right)+B_{i} \cos \left(k_{y} y\right). \label{eq:ezpar} 
\end{equation}

Using Eq. \ref{eq:ezpar}, $E_{y}=-\frac{i k_z}{k_{y}^{2}} \frac{\partial E_{z}}{\partial y}$ , and $H_{x}=\frac{i \omega \epsilon_0}{k_{y}^{2}} \frac{\partial E_{z}}{\partial y}$ one can derive $E_y$ and $H_x$ as well.
In addition, assuming that the Cartesian coordinates origin to be placed on the electron sheet, two sets of boundary conditions have been utilized to derive the dispersion relation (Eq. 3 in main text) as well as field distributions; first, the $z$-component of electric field is zero at PEC surfaces ($E_{z1} \arrowvert_{y=-d_1}=0$ and $E_{z2} \arrowvert_{y=d_2}=0$) and continuous at $y=0$ ($E_{z1} \arrowvert_{y=0^-}=E_{z2} \arrowvert_{y=0^+}$), second, at the place of electron beam ($y=0$), $H_z$ has a jump equal to the current density of electron beam Eq. \ref{eq:totaljap} times thickness of the electron beam $\delta$ ($H_{x1} \arrowvert_{y=0^-} - H_{x2} \arrowvert_{y=0^+}=J_z \delta$) (please note that for the parallel plate waveguide we have defined $n_0'$ which is $n_0=n_0'/\delta$, therefore when we multiply  $\delta$ with current density, $\delta$ will be cancelled out and $J_z\delta$ is independent of $\delta$). Hence, $e_y, e_z$ and $h_x$ in the two regions can be written as (Eq. \ref{eq:ez1}-\ref{eq:hx2}):

\begin{equation}
e_{z 1}= \cot{(k_{y1}d_1)} \sin \left(k_{y1} y\right)+ \cos \left(k_{y1} y\right), \label{eq:ez1} 
\end{equation}
\begin{equation}
e_{y 1}=-\frac{i k_{z}}{k_{y1}}( \cot{(k_{y1}d_1)} \cos \left(k_{y1} y\right)- \sin \left(k_{y1} y\right)), \label{eq:ey1} 
\end{equation}
\begin{equation}
h_{x 1}=\frac{i \omega \epsilon_0}{k_{y1}}( \cot{(k_{y1}d_1)} \cos \left(k_{y1} y\right)- \sin \left(k_{y1} y\right)), \label{eq:hx1} 
\end{equation}

\begin{equation}
e_{z 2}= -\cot{(k_{y2}d_2)} \sin \left(k_{y2} y\right)+ \cos \left(k_{y2} y\right), \label{eq:ez2} 
\end{equation}
\begin{equation}
e_{y 2}= \frac{i k_{z}}{k_{y2}}(\cot{(k_{y2}d_2)} \cos \left(k_{y2} y\right)+ \sin \left(k_{y2} y\right)), \label{eq:ey2} 
\end{equation}
\begin{equation}
h_{x 2}= -\frac{i \omega \epsilon_0}{k_{y2}}(\cot{(k_{y2}d_2)} \cos \left(k_{y2} y\right)+ \sin \left(k_{y2} y\right)), \label{eq:hx2} 
\end{equation}

where both regions have the same  permittivity therefore $k_{y1}=k_{y2}$.
\begin{figure}
    \centering
    \includegraphics[width=10cm]{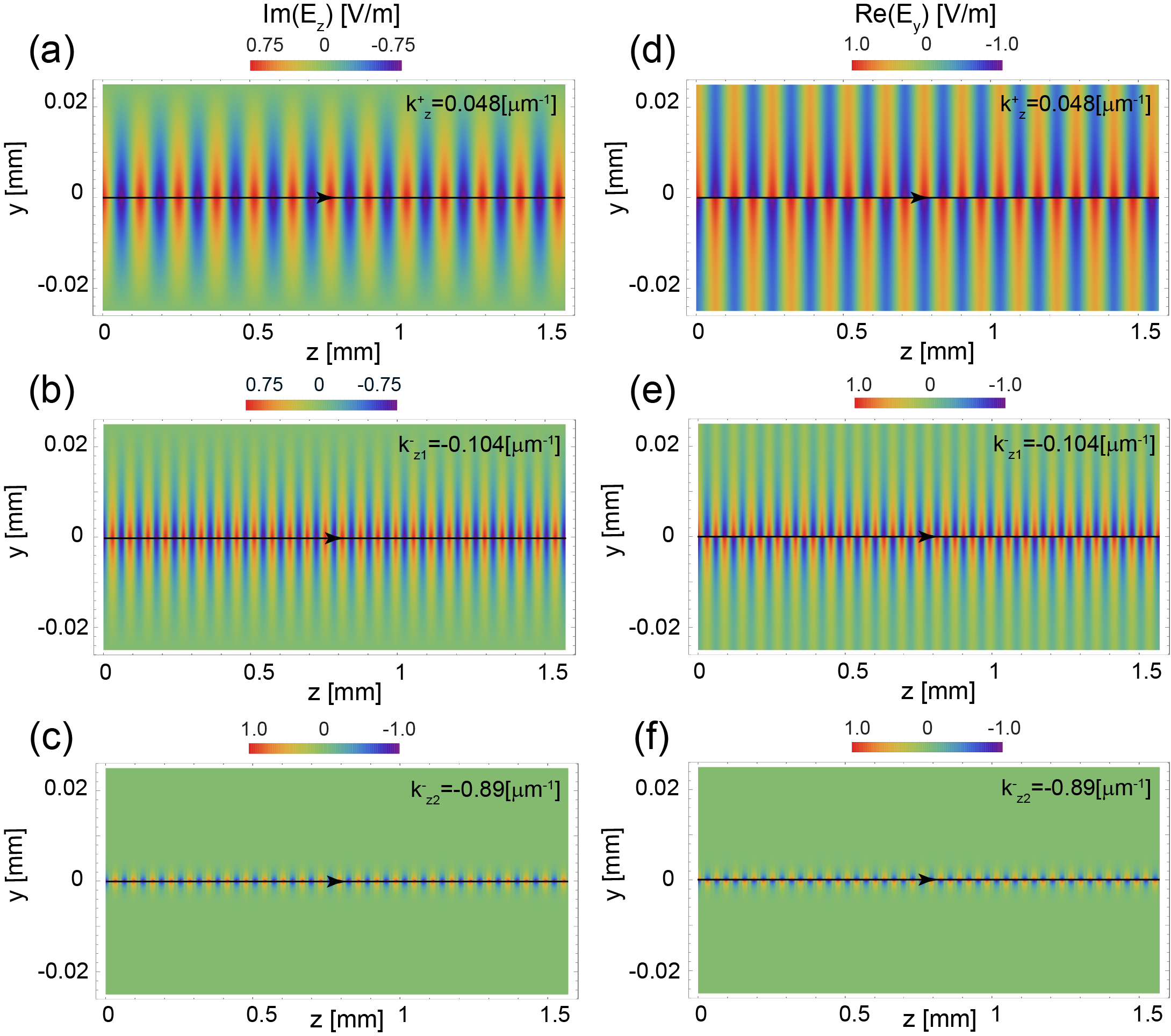}
    \caption{Time snapshots of the electric field profile of the $TM_{10}$ guided mode at $f=0.48$ $THz$ in the parallel plate waveguide with the electron beam having $v_0=c/30$ and $n_0=10^{17}$ $m^{-2}$. Panels (a),(b), and (c) depict the imaginary part of $E_z$ and (d), (e), and (f) show real part of $E_y$ for one $+z$ propagating and two $-z$ propagating waves with $k_z^+=0.048$ $\mu m^{-1}$, $k_{z1}^-=-0.104$ $\mu m^{-1}$, and  $k_{z2}^-=-0.89$ $\mu m^{-1}$, respectively. The black arrows sketch the electron motion with the tip of the arrow indicating the direction of electron flow.}
    \label{fig:fd}
\end{figure}

 Figure \ref{fig:fd} depicts time snapshots of the $y$- and $z$-components of the electric field in the $y$-$z$ plane for the $TM_{10}$ mode at $f=0.48$ $THz$, and in the presence of planar electron beam with velocity $v_0=c/30$, electron surface number density of $n_0'=10^{17}$ $m^{-2}$, and $d=0.05$ $mm$.
 
In the following, we derive the field distributions and dispersion relation of the cylindrical waveguide depicted in Fig. 3(a). The electric field distributions, considering $E_{z i}=e_{z i} e^{-i k_{z} z}$ where $i=1$ and $2$ in subscripts correspond to the inside and outside of the electron beam, respectively, are assumed to be of the following forms,

\begin{equation}
e_{z 1}=(A_1 \sin (n \varphi)+B_1 \cos (n \varphi)) J_{n}\left(k_{r1}r\right), \label{eq:ez1cyl}
\end{equation}
\begin{equation}
e_{z 2}=(A_1 \sin (n \varphi)+B_2 \cos (n \varphi))\left(C J_{n}\left(k_{r2} r\right)+D Y_{n}\left(k_{r2} r\right)\right), \label{eq:ez2cyl}
\end{equation}

where for $n=0$ ($TM_{10}$) and $B_1$, $B_2C$, $B_2D$ must be such that (i) at the surface of the PEC wall $E_{z2}=0$ and (ii) at the interface $r=b$ (the electron beam with vacuum), the Ampere's law must be satisfied ($\oint H. dl=\oint J.ds+i \omega \oint D.ds$) and $z$-components of the electric field are continuous ($E_{z1} \arrowvert_{r=b^-}=E_{z2} \arrowvert_{r=b^+}$). It is worth mentioning that, effective $zz$-component of the permittivity of the electron beam is assumed to be $\epsilon_e=1+\sigma/i \omega \epsilon_0$ along $z$-axis where $\sigma=J_z/E_z$. (the transverse components of the permittivity of the beam are not relevant here, since we assume that  the beam is very thin as compared to the wavelength.)  From these two boundary conditions and having $ k_r^2 \equiv k_0^2-k_z^2$,  Eq. 4 can be readily derived, and the field distributions in the limit of $b\ll \lambda$ are determined by, 

\begin{equation}
e_{z 1}=-\frac{Y_0(k_{r2}a)}{J_0(k_{r2}a)}J_0(k_{r2}b)+Y_0(k_{r2}b), \label{eq:ez1cylt}
\end{equation}
\begin{equation}
e_{r 1}=\frac{-i k_z}{2}(-\frac{Y_0(k_{r2}a)}{J_0(k_{r2}a)}J_0(k_{r2}b)+Y_0(k_{r2}b)) r, \label{eq:er1cylt}
\end{equation}
\begin{equation}
h_{\phi 1}=\frac{i \omega \epsilon_0 \epsilon_e}{2} (-\frac{Y_0(k_{r2}a)}{J_0(k_{r2}a)}J_0(k_{r2}b)+Y_0(k_{r2}b)) r, \label{eq:hp1cylt}
\end{equation}

\begin{equation}
e_{z 2}=-\frac{Y_0(k_{r2}a)}{J_0(k_{r2}a)}J_0(k_{r2}r)+Y_0(k_{r2}r), \label{eq:ez2cylt}
\end{equation}
\begin{equation}
e_{r 2}=\frac{-i k_z}{k_{r2}}(\frac{Y_0(k_{r2}a)}{J_0(k_{r2}a)}J_1(k_{r2}r)-Y_1(k_{r2}r)), \label{eq:er2cylt}
\end{equation}
\begin{equation}
h_{\phi 2}=-\frac{i \omega \epsilon_0 }{k_{r2}} (\frac{Y_0(k_{r2}a)}{J_0(k_{r2}a)}J_1(k_{r2}r)-Y_1(k_{r2}r) . \label{eq:hp2cylt}
\end{equation}

\begin{figure}
\includegraphics[width=10cm]{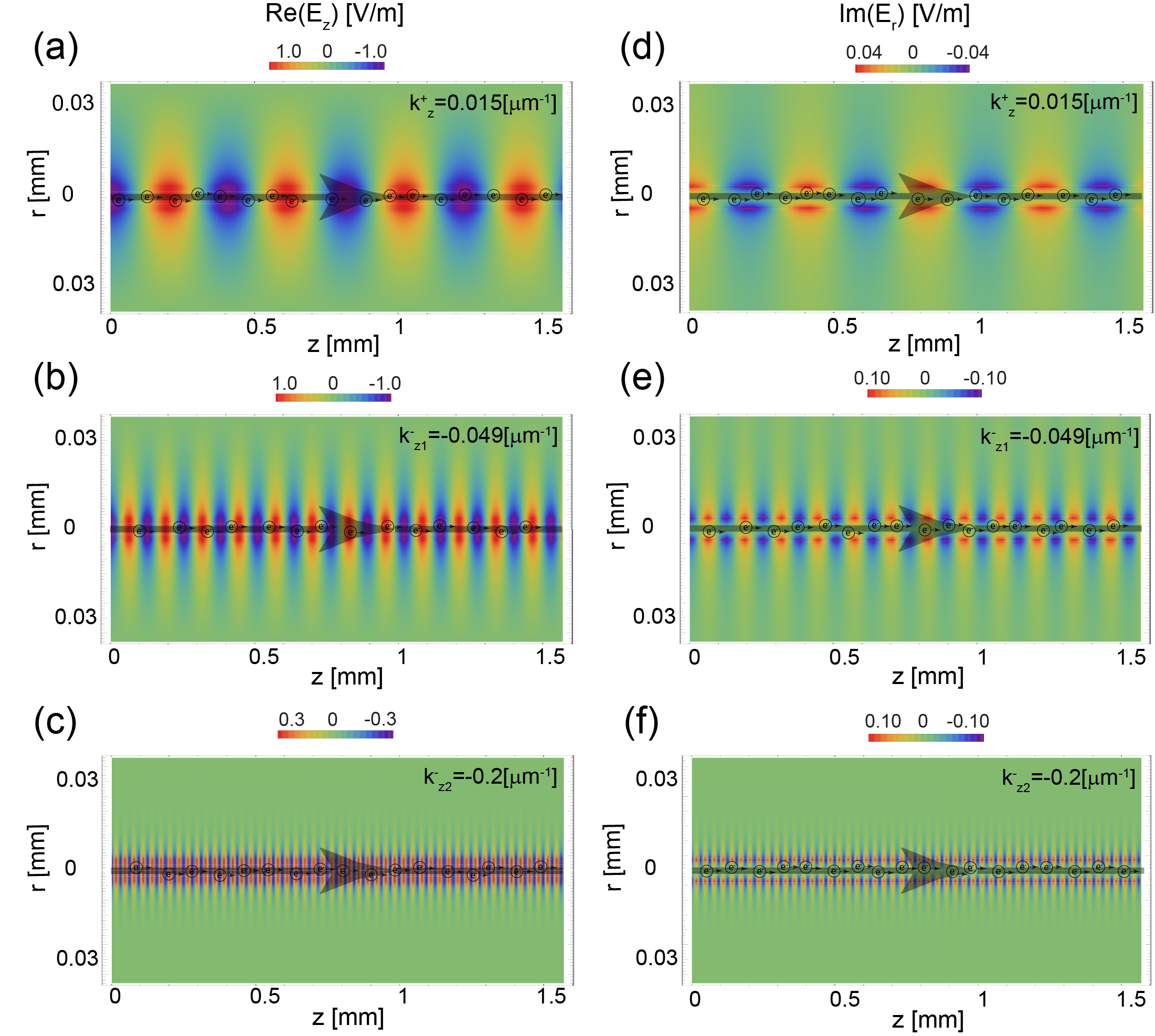}
\caption{\label{fig:fdcyl}Time snapshot of the electric field profile of the $TM_{10}$ guided wave at $f=0.08$ $THz$ in the circular-cylindrical waveguide with an axial electron pencil beam with $v_0=c/30$ and $n_0=10^{22}$ $m^{-3}$. Panels (a),(b), and (c) depict the real part of $E_z$ and (d), (e), and (f) present the imaginary part of $E_r$ for the $+z$ propagating and the two $-z$ propagating waves with $k_z^+=0.015$ $\mu m^{-1}$, $k_{z1}^-=-0.049$ $\mu m^{-1}$, and  $k_{z2}^-=-0.2$ $\mu m^{-1}$, respectively. The black arrows sketch the electron motion with the tip of the arrow indicating the direction of the electron flow.}
\end{figure}

Figure \ref{fig:fdcyl} shows the time snapshots of the $r$- and $z$-components of the electric field for this $TM_{10}$ mode when the electron velocity and electron density are taken equal to $v_0=c/30$ and $n_0=10^{22}$ $m^{-3}$, respectively.

\section{Numerical Simulation Methods}

In addition to our analytical investigation, we also conducted numerical simulations using the commercially available finite-element simulation software COMSOL Multiphysics\textsuperscript{\textregistered}. In order to take into account the nonlocal response of the electron beam in COMSOL, we used two
approaches: (1) for a given electron velocity, for each frequency of operation the conductivity of the electron beam has been swept as an independent parameter (treated as a constant independent of the wave vector) within a certain range of values, and for each value of the conductivity, we ran a simulation to obtain the corresponding value of the wavenumber $k_z$.  We then plot the calculated $k_z$ as a function of the conductivity.  We superimpose on this plot the analytical curve that gives the conductivity of the electron beam as a function of $k_z$ (Eq. 1). The intersection of the two curves yields the desired $k_z$ for the considered frequency. Repeating this procedure for many frequencies we obtain the dispersion diagram of the guided modes, taking fully into account the nonlocality, for various electron beam velocities; (2) in the second approach, we use a combination of analytical and numerical approaches, as follows. For a given electron velocity, we first use the analytical results of the dispersion relation to find the wavenumber for each frequency. Then using this wavenumber, we find the value of conductivity using the analytical expression of conductivity. We feed the calculated conductivity to COMSOL simulation and run a simulation to find the waves inside the waveguide for the given electron velocity and frequency. From this simulation, we can find the value of the wavenumber $k_z$ of the guided mode. This will provide us with the plot of $k_z$ as a function of frequency, for a given electron velocity. It is worth mentioning that, second method essentially serves to confirm analytical results. We explored both approaches and the numerical results shown in Figs. 1 and 2 are obtained using the second approach. Simulation results in Fig. 1(b) and Fig. 3(b) are shown with circles, clear demonstrating good agreement with analytical results shown using solid lines.

\begin{figure}
    \centering
      \includegraphics[width=10cm]{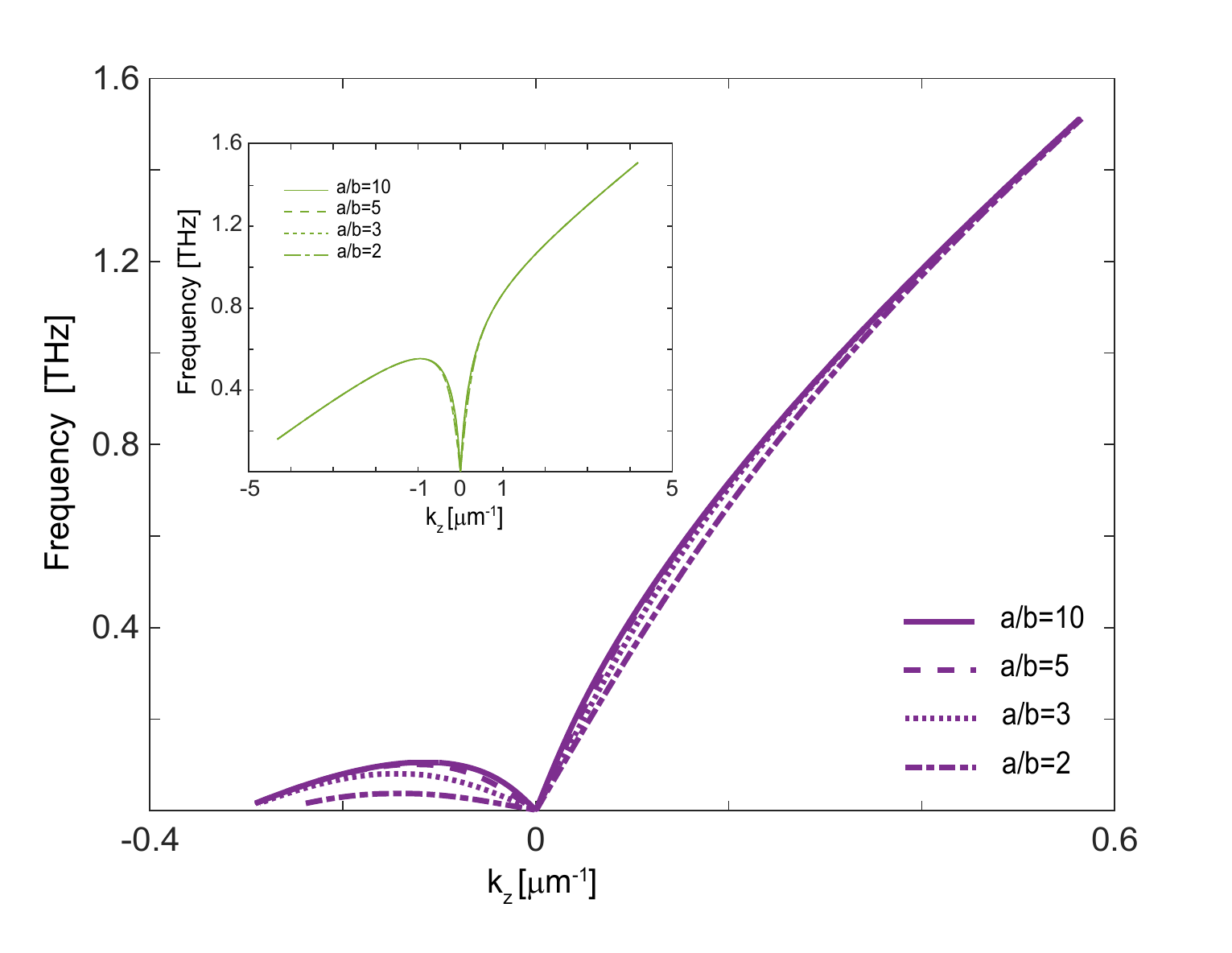}
    \caption{ Dispersion diagram of the cylindrical waveguide with various radii for $v_0=c/30$. The inset shows the dispersion diagram for a lower velocity electron beam assuming $v_0=c/300$. The electron beam radius is fixed at $3.8$ $um$.}
    \label{fig:c30300r}
\end{figure}  
\section{Influence of the outer Radius of the Cylindrical Waveguide}

The outer radius of the cylindrical waveguide does not have a significant influence on the dispersion diagram of the considered waveguides. For instance, in figure \ref{fig:c30300r}, we show the dispersion diagram of the cylindrical waveguide which was obtained using the analytical method for different radii $a/b=10,5,3,2$, considering $b=3.8$ $um$, $v_0=c/30$ and $v_0=c/300$. The results show that the variation of the outer radius is more perceptible for higher velocities.

%\begin{figure}[h]
%\includegraphics[width=17cm]{suppbib.png}
%\end{figure}

%% file: ms.bbl
\providecommand{\noopsort}[1]{}\providecommand{\singleletter}[1]{#1}%
\begin{thebibliography}{64}
\expandafter\ifx\csname natexlab\endcsname\relax\def\natexlab#1{#1}\fi
\expandafter\ifx\csname bibnamefont\endcsname\relax
  \def\bibnamefont#1{#1}\fi
\expandafter\ifx\csname bibfnamefont\endcsname\relax
  \def\bibfnamefont#1{#1}\fi
\expandafter\ifx\csname citenamefont\endcsname\relax
  \def\citenamefont#1{#1}\fi
\expandafter\ifx\csname url\endcsname\relax
  \def\url#1{\texttt{#1}}\fi
\expandafter\ifx\csname urlprefix\endcsname\relax\def\urlprefix{URL }\fi
\providecommand{\bibinfo}[2]{#2}
\providecommand{\eprint}[2][]{\url{#2}}

\bibitem[{\citenamefont{Strutt and Rayleigh}(1945)}]{rf1}
\bibinfo{author}{\bibfnamefont{J.~W.} \bibnamefont{Strutt}} \bibnamefont{and}
  \bibinfo{author}{\bibfnamefont{B.}~\bibnamefont{Rayleigh}},
  \emph{\bibinfo{title}{The theory of sound}} (\bibinfo{publisher}{Dover},
  \bibinfo{year}{1945}).

\bibitem[{\citenamefont{Lindell}(1992)}]{rf2}
\bibinfo{author}{\bibfnamefont{I.~V.} \bibnamefont{Lindell}}
  (\bibinfo{year}{1992}).

\bibitem[{\citenamefont{Staelin et~al.}(1994)\citenamefont{Staelin,
  Morgenthaler, and Kong}}]{rf3}
\bibinfo{author}{\bibfnamefont{D.~H.} \bibnamefont{Staelin}},
  \bibinfo{author}{\bibfnamefont{A.~W.} \bibnamefont{Morgenthaler}},
  \bibnamefont{and} \bibinfo{author}{\bibfnamefont{J.~A.} \bibnamefont{Kong}},
  \emph{\bibinfo{title}{Electromagnetic waves}} (\bibinfo{publisher}{Pearson
  Education India}, \bibinfo{year}{1994}).

\bibitem[{\citenamefont{De{\'a}k and F{\"u}l{\"o}p}(2012)}]{rf4}
\bibinfo{author}{\bibfnamefont{L.}~\bibnamefont{De{\'a}k}} \bibnamefont{and}
  \bibinfo{author}{\bibfnamefont{T.}~\bibnamefont{F{\"u}l{\"o}p}},
  \bibinfo{journal}{Annals of Physics} \textbf{\bibinfo{volume}{327}},
  \bibinfo{pages}{1050} (\bibinfo{year}{2012}).

\bibitem[{\citenamefont{Caloz et~al.}(2018)\citenamefont{Caloz, Al{\`u},
  Tretyakov, Sounas, Achouri, and Deck-L{\'e}ger}}]{rf5}
\bibinfo{author}{\bibfnamefont{C.}~\bibnamefont{Caloz}},
  \bibinfo{author}{\bibfnamefont{A.}~\bibnamefont{Al{\`u}}},
  \bibinfo{author}{\bibfnamefont{S.}~\bibnamefont{Tretyakov}},
  \bibinfo{author}{\bibfnamefont{D.}~\bibnamefont{Sounas}},
  \bibinfo{author}{\bibfnamefont{K.}~\bibnamefont{Achouri}}, \bibnamefont{and}
  \bibinfo{author}{\bibfnamefont{Z.-L.} \bibnamefont{Deck-L{\'e}ger}},
  \bibinfo{journal}{Physical Review Applied} \textbf{\bibinfo{volume}{10}},
  \bibinfo{pages}{047001} (\bibinfo{year}{2018}).

\bibitem[{\citenamefont{Gallo et~al.}(2001)\citenamefont{Gallo, Assanto,
  Parameswaran, and Fejer}}]{rf18}
\bibinfo{author}{\bibfnamefont{K.}~\bibnamefont{Gallo}},
  \bibinfo{author}{\bibfnamefont{G.}~\bibnamefont{Assanto}},
  \bibinfo{author}{\bibfnamefont{K.~R.} \bibnamefont{Parameswaran}},
  \bibnamefont{and} \bibinfo{author}{\bibfnamefont{M.~M.} \bibnamefont{Fejer}},
  \bibinfo{journal}{Applied Physics Letters} \textbf{\bibinfo{volume}{79}},
  \bibinfo{pages}{314} (\bibinfo{year}{2001}).

\bibitem[{\citenamefont{Solja{\v{c}}i{\'c}
  et~al.}(2003)\citenamefont{Solja{\v{c}}i{\'c}, Luo, Joannopoulos, and
  Fan}}]{rf29}
\bibinfo{author}{\bibfnamefont{M.}~\bibnamefont{Solja{\v{c}}i{\'c}}},
  \bibinfo{author}{\bibfnamefont{C.}~\bibnamefont{Luo}},
  \bibinfo{author}{\bibfnamefont{J.~D.} \bibnamefont{Joannopoulos}},
  \bibnamefont{and} \bibinfo{author}{\bibfnamefont{S.}~\bibnamefont{Fan}},
  \bibinfo{journal}{Optics letters} \textbf{\bibinfo{volume}{28}},
  \bibinfo{pages}{637} (\bibinfo{year}{2003}).

\bibitem[{\citenamefont{Yu et~al.}(2008)\citenamefont{Yu, Veronis, Wang, and
  Fan}}]{rf8}
\bibinfo{author}{\bibfnamefont{Z.}~\bibnamefont{Yu}},
  \bibinfo{author}{\bibfnamefont{G.}~\bibnamefont{Veronis}},
  \bibinfo{author}{\bibfnamefont{Z.}~\bibnamefont{Wang}}, \bibnamefont{and}
  \bibinfo{author}{\bibfnamefont{S.}~\bibnamefont{Fan}},
  \bibinfo{journal}{Physical review letters} \textbf{\bibinfo{volume}{100}},
  \bibinfo{pages}{023902} (\bibinfo{year}{2008}).

\bibitem[{\citenamefont{Yu and Fan}(2009)}]{rf6}
\bibinfo{author}{\bibfnamefont{Z.}~\bibnamefont{Yu}} \bibnamefont{and}
  \bibinfo{author}{\bibfnamefont{S.}~\bibnamefont{Fan}},
  \bibinfo{journal}{Nature photonics} \textbf{\bibinfo{volume}{3}},
  \bibinfo{pages}{91} (\bibinfo{year}{2009}).

\bibitem[{\citenamefont{Ramezani et~al.}(2010)\citenamefont{Ramezani, Kottos,
  El-Ganainy, and Christodoulides}}]{rf30}
\bibinfo{author}{\bibfnamefont{H.}~\bibnamefont{Ramezani}},
  \bibinfo{author}{\bibfnamefont{T.}~\bibnamefont{Kottos}},
  \bibinfo{author}{\bibfnamefont{R.}~\bibnamefont{El-Ganainy}},
  \bibnamefont{and} \bibinfo{author}{\bibfnamefont{D.~N.}
  \bibnamefont{Christodoulides}}, \bibinfo{journal}{Physical Review A}
  \textbf{\bibinfo{volume}{82}}, \bibinfo{pages}{043803}
  (\bibinfo{year}{2010}).

\bibitem[{\citenamefont{Kodera et~al.}(2011{\natexlab{a}})\citenamefont{Kodera,
  Sounas, and Caloz}}]{rf27}
\bibinfo{author}{\bibfnamefont{T.}~\bibnamefont{Kodera}},
  \bibinfo{author}{\bibfnamefont{D.~L.} \bibnamefont{Sounas}},
  \bibnamefont{and} \bibinfo{author}{\bibfnamefont{C.}~\bibnamefont{Caloz}},
  \bibinfo{journal}{IEEE Antennas and Wireless Propagation Letters}
  \textbf{\bibinfo{volume}{10}}, \bibinfo{pages}{1551}
  (\bibinfo{year}{2011}{\natexlab{a}}).

\bibitem[{\citenamefont{Kodera et~al.}(2011{\natexlab{b}})\citenamefont{Kodera,
  Sounas, and Caloz}}]{rf25}
\bibinfo{author}{\bibfnamefont{T.}~\bibnamefont{Kodera}},
  \bibinfo{author}{\bibfnamefont{D.~L.} \bibnamefont{Sounas}},
  \bibnamefont{and} \bibinfo{author}{\bibfnamefont{C.}~\bibnamefont{Caloz}},
  \bibinfo{journal}{Applied Physics Letters} \textbf{\bibinfo{volume}{99}},
  \bibinfo{pages}{031114} (\bibinfo{year}{2011}{\natexlab{b}}).

\bibitem[{\citenamefont{Shadrivov et~al.}(2011)\citenamefont{Shadrivov,
  Fedotov, Powell, Kivshar, and Zheludev}}]{rf19}
\bibinfo{author}{\bibfnamefont{I.~V.} \bibnamefont{Shadrivov}},
  \bibinfo{author}{\bibfnamefont{V.~A.} \bibnamefont{Fedotov}},
  \bibinfo{author}{\bibfnamefont{D.~A.} \bibnamefont{Powell}},
  \bibinfo{author}{\bibfnamefont{Y.~S.} \bibnamefont{Kivshar}},
  \bibnamefont{and} \bibinfo{author}{\bibfnamefont{N.~I.}
  \bibnamefont{Zheludev}}, \bibinfo{journal}{New Journal of Physics}
  \textbf{\bibinfo{volume}{13}}, \bibinfo{pages}{033025}
  (\bibinfo{year}{2011}).

\bibitem[{\citenamefont{Bi et~al.}(2011)\citenamefont{Bi, Hu, Jiang, Kim,
  Dionne, Kimerling, and Ross}}]{rf13}
\bibinfo{author}{\bibfnamefont{L.}~\bibnamefont{Bi}},
  \bibinfo{author}{\bibfnamefont{J.}~\bibnamefont{Hu}},
  \bibinfo{author}{\bibfnamefont{P.}~\bibnamefont{Jiang}},
  \bibinfo{author}{\bibfnamefont{D.~H.} \bibnamefont{Kim}},
  \bibinfo{author}{\bibfnamefont{G.~F.} \bibnamefont{Dionne}},
  \bibinfo{author}{\bibfnamefont{L.~C.} \bibnamefont{Kimerling}},
  \bibnamefont{and} \bibinfo{author}{\bibfnamefont{C.}~\bibnamefont{Ross}},
  \bibinfo{journal}{Nature Photonics} \textbf{\bibinfo{volume}{5}},
  \bibinfo{pages}{758} (\bibinfo{year}{2011}).

\bibitem[{\citenamefont{Lira et~al.}(2012)\citenamefont{Lira, Yu, Fan, and
  Lipson}}]{rf11}
\bibinfo{author}{\bibfnamefont{H.}~\bibnamefont{Lira}},
  \bibinfo{author}{\bibfnamefont{Z.}~\bibnamefont{Yu}},
  \bibinfo{author}{\bibfnamefont{S.}~\bibnamefont{Fan}}, \bibnamefont{and}
  \bibinfo{author}{\bibfnamefont{M.}~\bibnamefont{Lipson}},
  \bibinfo{journal}{Physical review letters} \textbf{\bibinfo{volume}{109}},
  \bibinfo{pages}{033901} (\bibinfo{year}{2012}).

\bibitem[{\citenamefont{Fan et~al.}(2012)\citenamefont{Fan, Wang, Varghese,
  Shen, Niu, Xuan, Weiner, and Qi}}]{rf15}
\bibinfo{author}{\bibfnamefont{L.}~\bibnamefont{Fan}},
  \bibinfo{author}{\bibfnamefont{J.}~\bibnamefont{Wang}},
  \bibinfo{author}{\bibfnamefont{L.~T.} \bibnamefont{Varghese}},
  \bibinfo{author}{\bibfnamefont{H.}~\bibnamefont{Shen}},
  \bibinfo{author}{\bibfnamefont{B.}~\bibnamefont{Niu}},
  \bibinfo{author}{\bibfnamefont{Y.}~\bibnamefont{Xuan}},
  \bibinfo{author}{\bibfnamefont{A.~M.} \bibnamefont{Weiner}},
  \bibnamefont{and} \bibinfo{author}{\bibfnamefont{M.}~\bibnamefont{Qi}},
  \bibinfo{journal}{Science} \textbf{\bibinfo{volume}{335}},
  \bibinfo{pages}{447} (\bibinfo{year}{2012}).

\bibitem[{\citenamefont{Luukkonen et~al.}(2012)\citenamefont{Luukkonen,
  Chettiar, and Engheta}}]{rf21}
\bibinfo{author}{\bibfnamefont{O.}~\bibnamefont{Luukkonen}},
  \bibinfo{author}{\bibfnamefont{U.~K.} \bibnamefont{Chettiar}},
  \bibnamefont{and} \bibinfo{author}{\bibfnamefont{N.}~\bibnamefont{Engheta}},
  \bibinfo{journal}{IEEE Antennas and Wireless Propagation Letters}
  \textbf{\bibinfo{volume}{11}}, \bibinfo{pages}{1398} (\bibinfo{year}{2012}).

\bibitem[{\citenamefont{Wang et~al.}(2012)\citenamefont{Wang, Wang, Wang,
  Zhang, Huangfu, Joannopoulos, Solja{\v{c}}i{\'c}, and Ran}}]{rf26}
\bibinfo{author}{\bibfnamefont{Z.}~\bibnamefont{Wang}},
  \bibinfo{author}{\bibfnamefont{Z.}~\bibnamefont{Wang}},
  \bibinfo{author}{\bibfnamefont{J.}~\bibnamefont{Wang}},
  \bibinfo{author}{\bibfnamefont{B.}~\bibnamefont{Zhang}},
  \bibinfo{author}{\bibfnamefont{J.}~\bibnamefont{Huangfu}},
  \bibinfo{author}{\bibfnamefont{J.~D.} \bibnamefont{Joannopoulos}},
  \bibinfo{author}{\bibfnamefont{M.}~\bibnamefont{Solja{\v{c}}i{\'c}}},
  \bibnamefont{and} \bibinfo{author}{\bibfnamefont{L.}~\bibnamefont{Ran}},
  \bibinfo{journal}{Proceedings of the National Academy of Sciences}
  \textbf{\bibinfo{volume}{109}}, \bibinfo{pages}{13194}
  (\bibinfo{year}{2012}).

\bibitem[{\citenamefont{Sounas et~al.}(2013)\citenamefont{Sounas, Caloz, and
  Al{\`u}}}]{rf24}
\bibinfo{author}{\bibfnamefont{D.~L.} \bibnamefont{Sounas}},
  \bibinfo{author}{\bibfnamefont{C.}~\bibnamefont{Caloz}}, \bibnamefont{and}
  \bibinfo{author}{\bibfnamefont{A.}~\bibnamefont{Al{\`u}}},
  \bibinfo{journal}{Nature communications} \textbf{\bibinfo{volume}{4}},
  \bibinfo{pages}{2407} (\bibinfo{year}{2013}).

\bibitem[{\citenamefont{Davoyan and Engheta}(2014)}]{davoyan2014electrically}
\bibinfo{author}{\bibfnamefont{A.}~\bibnamefont{Davoyan}} \bibnamefont{and}
  \bibinfo{author}{\bibfnamefont{N.}~\bibnamefont{Engheta}},
  \bibinfo{journal}{Nature communications} \textbf{\bibinfo{volume}{5}},
  \bibinfo{pages}{1} (\bibinfo{year}{2014}).

\bibitem[{\citenamefont{Fleury et~al.}(2014)\citenamefont{Fleury, Sounas,
  Sieck, Haberman, and Al{\`u}}}]{rf7}
\bibinfo{author}{\bibfnamefont{R.}~\bibnamefont{Fleury}},
  \bibinfo{author}{\bibfnamefont{D.~L.} \bibnamefont{Sounas}},
  \bibinfo{author}{\bibfnamefont{C.~F.} \bibnamefont{Sieck}},
  \bibinfo{author}{\bibfnamefont{M.~R.} \bibnamefont{Haberman}},
  \bibnamefont{and} \bibinfo{author}{\bibfnamefont{A.}~\bibnamefont{Al{\`u}}},
  \bibinfo{journal}{Science} \textbf{\bibinfo{volume}{343}},
  \bibinfo{pages}{516} (\bibinfo{year}{2014}).

\bibitem[{\citenamefont{Estep et~al.}(2014)\citenamefont{Estep, Sounas, Soric,
  and Al{\`u}}}]{rf10}
\bibinfo{author}{\bibfnamefont{N.~A.} \bibnamefont{Estep}},
  \bibinfo{author}{\bibfnamefont{D.~L.} \bibnamefont{Sounas}},
  \bibinfo{author}{\bibfnamefont{J.}~\bibnamefont{Soric}}, \bibnamefont{and}
  \bibinfo{author}{\bibfnamefont{A.}~\bibnamefont{Al{\`u}}},
  \bibinfo{journal}{Nature Physics} \textbf{\bibinfo{volume}{10}},
  \bibinfo{pages}{923} (\bibinfo{year}{2014}).

\bibitem[{\citenamefont{Sounas and Al{\`u}̀}(2014)}]{rf14}
\bibinfo{author}{\bibfnamefont{D.~L.} \bibnamefont{Sounas}} \bibnamefont{and}
  \bibinfo{author}{\bibfnamefont{A.}~\bibnamefont{Al{\`u}̀}},
  \bibinfo{journal}{ACS photonics} \textbf{\bibinfo{volume}{1}},
  \bibinfo{pages}{198} (\bibinfo{year}{2014}).

\bibitem[{\citenamefont{Mahmoud et~al.}(2015)\citenamefont{Mahmoud, Davoyan,
  and Engheta}}]{rf16}
\bibinfo{author}{\bibfnamefont{A.~M.} \bibnamefont{Mahmoud}},
  \bibinfo{author}{\bibfnamefont{A.~R.} \bibnamefont{Davoyan}},
  \bibnamefont{and} \bibinfo{author}{\bibfnamefont{N.}~\bibnamefont{Engheta}},
  \bibinfo{journal}{Nature communications} \textbf{\bibinfo{volume}{6}},
  \bibinfo{pages}{1} (\bibinfo{year}{2015}).

\bibitem[{\citenamefont{Sounas and Al{\`u}}(2017{\natexlab{a}})}]{rf12}
\bibinfo{author}{\bibfnamefont{D.~L.} \bibnamefont{Sounas}} \bibnamefont{and}
  \bibinfo{author}{\bibfnamefont{A.}~\bibnamefont{Al{\`u}}},
  \bibinfo{journal}{Nature Photonics} \textbf{\bibinfo{volume}{11}},
  \bibinfo{pages}{774} (\bibinfo{year}{2017}{\natexlab{a}}).

\bibitem[{\citenamefont{Sounas and Al{\`u}}(2018{\natexlab{a}})}]{rf17}
\bibinfo{author}{\bibfnamefont{D.~L.} \bibnamefont{Sounas}} \bibnamefont{and}
  \bibinfo{author}{\bibfnamefont{A.}~\bibnamefont{Al{\`u}}},
  \bibinfo{journal}{IEEE Antennas and Wireless Propagation Letters}
  \textbf{\bibinfo{volume}{17}}, \bibinfo{pages}{1958}
  (\bibinfo{year}{2018}{\natexlab{a}}).

\bibitem[{\citenamefont{Bliokh et~al.}(2018)\citenamefont{Bliokh,
  Rodr{\'\i}guez-Fortu{\~n}o, Bekshaev, Kivshar, and
  Nori}}]{bliokh2018electric}
\bibinfo{author}{\bibfnamefont{K.}~\bibnamefont{Bliokh}},
  \bibinfo{author}{\bibfnamefont{F.~J.}
  \bibnamefont{Rodr{\'\i}guez-Fortu{\~n}o}},
  \bibinfo{author}{\bibfnamefont{A.}~\bibnamefont{Bekshaev}},
  \bibinfo{author}{\bibfnamefont{Y.}~\bibnamefont{Kivshar}}, \bibnamefont{and}
  \bibinfo{author}{\bibfnamefont{F.}~\bibnamefont{Nori}},
  \bibinfo{journal}{Optics letters} \textbf{\bibinfo{volume}{43}},
  \bibinfo{pages}{963} (\bibinfo{year}{2018}).

\bibitem[{\citenamefont{Mazor and Al{\`u}}(2019)}]{rf28}
\bibinfo{author}{\bibfnamefont{Y.}~\bibnamefont{Mazor}} \bibnamefont{and}
  \bibinfo{author}{\bibfnamefont{A.}~\bibnamefont{Al{\`u}}},
  \bibinfo{journal}{Physical Review B} \textbf{\bibinfo{volume}{99}},
  \bibinfo{pages}{045407} (\bibinfo{year}{2019}).

\bibitem[{\citenamefont{Li et~al.}(2019)\citenamefont{Li, Peng, Han, Miri, Li,
  Xiao, Zhu, Zhao, Al{\`u}, Fan et~al.}}]{rf23}
\bibinfo{author}{\bibfnamefont{Y.}~\bibnamefont{Li}},
  \bibinfo{author}{\bibfnamefont{Y.-G.} \bibnamefont{Peng}},
  \bibinfo{author}{\bibfnamefont{L.}~\bibnamefont{Han}},
  \bibinfo{author}{\bibfnamefont{M.-A.} \bibnamefont{Miri}},
  \bibinfo{author}{\bibfnamefont{W.}~\bibnamefont{Li}},
  \bibinfo{author}{\bibfnamefont{M.}~\bibnamefont{Xiao}},
  \bibinfo{author}{\bibfnamefont{X.-F.} \bibnamefont{Zhu}},
  \bibinfo{author}{\bibfnamefont{J.}~\bibnamefont{Zhao}},
  \bibinfo{author}{\bibfnamefont{A.}~\bibnamefont{Al{\`u}}},
  \bibinfo{author}{\bibfnamefont{S.}~\bibnamefont{Fan}}, \bibnamefont{et~al.},
  \bibinfo{journal}{Science} \textbf{\bibinfo{volume}{364}},
  \bibinfo{pages}{170} (\bibinfo{year}{2019}).

\bibitem[{\citenamefont{Mann et~al.}(2019)\citenamefont{Mann, Sounas, and
  Al{\`u}}}]{rf22}
\bibinfo{author}{\bibfnamefont{S.~A.} \bibnamefont{Mann}},
  \bibinfo{author}{\bibfnamefont{D.~L.} \bibnamefont{Sounas}},
  \bibnamefont{and} \bibinfo{author}{\bibfnamefont{A.}~\bibnamefont{Al{\`u}}},
  \bibinfo{journal}{Physical Review B} \textbf{\bibinfo{volume}{100}},
  \bibinfo{pages}{020303} (\bibinfo{year}{2019}).

\bibitem[{\citenamefont{Fernandes and Silveirinha}(2019)}]{rf20}
\bibinfo{author}{\bibfnamefont{D.~E.} \bibnamefont{Fernandes}}
  \bibnamefont{and} \bibinfo{author}{\bibfnamefont{M.~G.}
  \bibnamefont{Silveirinha}}, \bibinfo{journal}{Physical Review Applied}
  \textbf{\bibinfo{volume}{12}}, \bibinfo{pages}{014021}
  (\bibinfo{year}{2019}).

\bibitem[{\citenamefont{Mock et~al.}(2019)\citenamefont{Mock, Sounas, and
  Al{\`u}}}]{rf9}
\bibinfo{author}{\bibfnamefont{A.}~\bibnamefont{Mock}},
  \bibinfo{author}{\bibfnamefont{D.}~\bibnamefont{Sounas}}, \bibnamefont{and}
  \bibinfo{author}{\bibfnamefont{A.}~\bibnamefont{Al{\`u}}},
  \bibinfo{journal}{ACS Photonics}  (\bibinfo{year}{2019}).

\bibitem[{\citenamefont{Lax et~al.}(1954)\citenamefont{Lax, Button, and
  Roth}}]{rf31}
\bibinfo{author}{\bibfnamefont{B.}~\bibnamefont{Lax}},
  \bibinfo{author}{\bibfnamefont{K.~J.} \bibnamefont{Button}},
  \bibnamefont{and} \bibinfo{author}{\bibfnamefont{L.~M.} \bibnamefont{Roth}},
  \bibinfo{journal}{Journal of Applied Physics} \textbf{\bibinfo{volume}{25}},
  \bibinfo{pages}{1413} (\bibinfo{year}{1954}).

\bibitem[{\citenamefont{Fox et~al.}(1955)\citenamefont{Fox, Miller, and
  Weiss}}]{rf32}
\bibinfo{author}{\bibfnamefont{A.}~\bibnamefont{Fox}},
  \bibinfo{author}{\bibfnamefont{S.}~\bibnamefont{Miller}}, \bibnamefont{and}
  \bibinfo{author}{\bibfnamefont{M.}~\bibnamefont{Weiss}},
  \bibinfo{journal}{Bell System Technical Journal}
  \textbf{\bibinfo{volume}{34}}, \bibinfo{pages}{5} (\bibinfo{year}{1955}).

\bibitem[{\citenamefont{Button and Lax}(1956)}]{rf33}
\bibinfo{author}{\bibfnamefont{K.}~\bibnamefont{Button}} \bibnamefont{and}
  \bibinfo{author}{\bibfnamefont{B.}~\bibnamefont{Lax}}, \bibinfo{journal}{IRE
  Transactions on Antennas and Propagation} \textbf{\bibinfo{volume}{4}},
  \bibinfo{pages}{531} (\bibinfo{year}{1956}).

\bibitem[{\citenamefont{Brion et~al.}(1972)\citenamefont{Brion, Wallis,
  Hartstein, and Burstein}}]{rf34}
\bibinfo{author}{\bibfnamefont{J.}~\bibnamefont{Brion}},
  \bibinfo{author}{\bibfnamefont{R.}~\bibnamefont{Wallis}},
  \bibinfo{author}{\bibfnamefont{A.}~\bibnamefont{Hartstein}},
  \bibnamefont{and} \bibinfo{author}{\bibfnamefont{E.}~\bibnamefont{Burstein}},
  \bibinfo{journal}{Physical Review Letters} \textbf{\bibinfo{volume}{28}},
  \bibinfo{pages}{1455} (\bibinfo{year}{1972}).

\bibitem[{\citenamefont{Furlani}(2001)}]{rf35}
\bibinfo{author}{\bibfnamefont{E.~P.} \bibnamefont{Furlani}},
  \emph{\bibinfo{title}{Permanent magnet and electromechanical devices:
  materials, analysis, and applications}} (\bibinfo{publisher}{Academic press},
  \bibinfo{year}{2001}).

\bibitem[{\citenamefont{Lannebere and Silveirinha}(2016)}]{rf36}
\bibinfo{author}{\bibfnamefont{S.}~\bibnamefont{Lannebere}} \bibnamefont{and}
  \bibinfo{author}{\bibfnamefont{M.~G.} \bibnamefont{Silveirinha}},
  \bibinfo{journal}{Physical Review A} \textbf{\bibinfo{volume}{94}},
  \bibinfo{pages}{033810} (\bibinfo{year}{2016}).

\bibitem[{\citenamefont{Weis et~al.}(2010)\citenamefont{Weis, Rivi{\`e}re,
  Del{\'e}glise, Gavartin, Arcizet, Schliesser, and Kippenberg}}]{rf37}
\bibinfo{author}{\bibfnamefont{S.}~\bibnamefont{Weis}},
  \bibinfo{author}{\bibfnamefont{R.}~\bibnamefont{Rivi{\`e}re}},
  \bibinfo{author}{\bibfnamefont{S.}~\bibnamefont{Del{\'e}glise}},
  \bibinfo{author}{\bibfnamefont{E.}~\bibnamefont{Gavartin}},
  \bibinfo{author}{\bibfnamefont{O.}~\bibnamefont{Arcizet}},
  \bibinfo{author}{\bibfnamefont{A.}~\bibnamefont{Schliesser}},
  \bibnamefont{and} \bibinfo{author}{\bibfnamefont{T.~J.}
  \bibnamefont{Kippenberg}}, \bibinfo{journal}{Science}
  \textbf{\bibinfo{volume}{330}}, \bibinfo{pages}{1520} (\bibinfo{year}{2010}).

\bibitem[{\citenamefont{Hafezi and Rabl}(2012)}]{rf38}
\bibinfo{author}{\bibfnamefont{M.}~\bibnamefont{Hafezi}} \bibnamefont{and}
  \bibinfo{author}{\bibfnamefont{P.}~\bibnamefont{Rabl}},
  \bibinfo{journal}{Optics express} \textbf{\bibinfo{volume}{20}},
  \bibinfo{pages}{7672} (\bibinfo{year}{2012}).

\bibitem[{\citenamefont{Huidobro et~al.}(2019)\citenamefont{Huidobro, Galiffi,
  Guenneau, Craster, and Pendry}}]{rf24a}
\bibinfo{author}{\bibfnamefont{P.~A.} \bibnamefont{Huidobro}},
  \bibinfo{author}{\bibfnamefont{E.}~\bibnamefont{Galiffi}},
  \bibinfo{author}{\bibfnamefont{S.}~\bibnamefont{Guenneau}},
  \bibinfo{author}{\bibfnamefont{R.~V.} \bibnamefont{Craster}},
  \bibnamefont{and} \bibinfo{author}{\bibfnamefont{J.}~\bibnamefont{Pendry}},
  \bibinfo{journal}{Proceedings of the National Academy of Sciences}
  \textbf{\bibinfo{volume}{116}}, \bibinfo{pages}{24943}
  (\bibinfo{year}{2019}).

\bibitem[{\citenamefont{Shi et~al.}(2015)\citenamefont{Shi, Yu, and
  Fan}}]{rf39}
\bibinfo{author}{\bibfnamefont{Y.}~\bibnamefont{Shi}},
  \bibinfo{author}{\bibfnamefont{Z.}~\bibnamefont{Yu}}, \bibnamefont{and}
  \bibinfo{author}{\bibfnamefont{S.}~\bibnamefont{Fan}},
  \bibinfo{journal}{Nature photonics} \textbf{\bibinfo{volume}{9}},
  \bibinfo{pages}{388} (\bibinfo{year}{2015}).

\bibitem[{\citenamefont{Sounas and Al{\`u}}(2018{\natexlab{b}})}]{rf39a}
\bibinfo{author}{\bibfnamefont{D.~L.} \bibnamefont{Sounas}} \bibnamefont{and}
  \bibinfo{author}{\bibfnamefont{A.}~\bibnamefont{Al{\`u}}},
  \bibinfo{journal}{Physical Review B} \textbf{\bibinfo{volume}{97}},
  \bibinfo{pages}{115431} (\bibinfo{year}{2018}{\natexlab{b}}).

\bibitem[{\citenamefont{Sounas and Al{\`u}}(2017{\natexlab{b}})}]{rf39b}
\bibinfo{author}{\bibfnamefont{D.~L.} \bibnamefont{Sounas}} \bibnamefont{and}
  \bibinfo{author}{\bibfnamefont{A.}~\bibnamefont{Al{\`u}}},
  \bibinfo{journal}{Physical review letters} \textbf{\bibinfo{volume}{118}},
  \bibinfo{pages}{154302} (\bibinfo{year}{2017}{\natexlab{b}}).

\bibitem[{\citenamefont{Morgado and Silveirinha}(2018)}]{rf40}
\bibinfo{author}{\bibfnamefont{T.~A.} \bibnamefont{Morgado}} \bibnamefont{and}
  \bibinfo{author}{\bibfnamefont{M.~G.} \bibnamefont{Silveirinha}},
  \bibinfo{journal}{ACS Photonics} \textbf{\bibinfo{volume}{5}},
  \bibinfo{pages}{4253} (\bibinfo{year}{2018}).

\bibitem[{\citenamefont{Correas-Serrano and Gomez-Diaz}(2019)}]{rf42g}
\bibinfo{author}{\bibfnamefont{D.}~\bibnamefont{Correas-Serrano}}
  \bibnamefont{and}
  \bibinfo{author}{\bibfnamefont{J.}~\bibnamefont{Gomez-Diaz}},
  \bibinfo{journal}{Physical Review B} \textbf{\bibinfo{volume}{100}},
  \bibinfo{pages}{081410} (\bibinfo{year}{2019}).

\bibitem[{\citenamefont{Morgado and Silveirinha}(2017)}]{rf41}
\bibinfo{author}{\bibfnamefont{T.~A.} \bibnamefont{Morgado}} \bibnamefont{and}
  \bibinfo{author}{\bibfnamefont{M.~G.} \bibnamefont{Silveirinha}},
  \bibinfo{journal}{Physical review letters} \textbf{\bibinfo{volume}{119}},
  \bibinfo{pages}{133901} (\bibinfo{year}{2017}).

\bibitem[{\citenamefont{Borgnia et~al.}(2015)\citenamefont{Borgnia, Phan, and
  Levitov}}]{rf42}
\bibinfo{author}{\bibfnamefont{D.~S.} \bibnamefont{Borgnia}},
  \bibinfo{author}{\bibfnamefont{T.~V.} \bibnamefont{Phan}}, \bibnamefont{and}
  \bibinfo{author}{\bibfnamefont{L.~S.} \bibnamefont{Levitov}},
  \bibinfo{journal}{arXiv preprint arXiv:1512.09044}  (\bibinfo{year}{2015}).

\bibitem[{\citenamefont{Van~Duppen et~al.}(2016)\citenamefont{Van~Duppen,
  Tomadin, Grigorenko, and Polini}}]{rf42a}
\bibinfo{author}{\bibfnamefont{B.}~\bibnamefont{Van~Duppen}},
  \bibinfo{author}{\bibfnamefont{A.}~\bibnamefont{Tomadin}},
  \bibinfo{author}{\bibfnamefont{A.~N.} \bibnamefont{Grigorenko}},
  \bibnamefont{and} \bibinfo{author}{\bibfnamefont{M.}~\bibnamefont{Polini}},
  \bibinfo{journal}{2D Materials} \textbf{\bibinfo{volume}{3}},
  \bibinfo{pages}{015011} (\bibinfo{year}{2016}).

\bibitem[{\citenamefont{Sydoruk et~al.}(2012)\citenamefont{Sydoruk, Syms, and
  Solymar}}]{rf44}
\bibinfo{author}{\bibfnamefont{O.}~\bibnamefont{Sydoruk}},
  \bibinfo{author}{\bibfnamefont{R.}~\bibnamefont{Syms}}, \bibnamefont{and}
  \bibinfo{author}{\bibfnamefont{L.}~\bibnamefont{Solymar}},
  \bibinfo{journal}{Journal of Applied Physics} \textbf{\bibinfo{volume}{112}},
  \bibinfo{pages}{104512} (\bibinfo{year}{2012}).

\bibitem[{\citenamefont{Dorgan et~al.}(2010)\citenamefont{Dorgan, Bae, and
  Pop}}]{rf43}
\bibinfo{author}{\bibfnamefont{V.~E.} \bibnamefont{Dorgan}},
  \bibinfo{author}{\bibfnamefont{M.-H.} \bibnamefont{Bae}}, \bibnamefont{and}
  \bibinfo{author}{\bibfnamefont{E.}~\bibnamefont{Pop}},
  \bibinfo{journal}{Applied Physics Letters} \textbf{\bibinfo{volume}{97}},
  \bibinfo{pages}{082112} (\bibinfo{year}{2010}).

\bibitem[{\citenamefont{Barnard et~al.}(2011)\citenamefont{Barnard, Coenen,
  Vesseur, Polman, and Brongersma}}]{CLparam}
\bibinfo{author}{\bibfnamefont{E.~S.} \bibnamefont{Barnard}},
  \bibinfo{author}{\bibfnamefont{T.}~\bibnamefont{Coenen}},
  \bibinfo{author}{\bibfnamefont{E.~J.~R.} \bibnamefont{Vesseur}},
  \bibinfo{author}{\bibfnamefont{A.}~\bibnamefont{Polman}}, \bibnamefont{and}
  \bibinfo{author}{\bibfnamefont{M.~L.} \bibnamefont{Brongersma}},
  \bibinfo{journal}{Nano letters} \textbf{\bibinfo{volume}{11}},
  \bibinfo{pages}{4265} (\bibinfo{year}{2011}).

\bibitem[{\citenamefont{Polman et~al.}(2019)\citenamefont{Polman, Kociak, and
  de~Abajo}}]{polman2019electron}
\bibinfo{author}{\bibfnamefont{A.}~\bibnamefont{Polman}},
  \bibinfo{author}{\bibfnamefont{M.}~\bibnamefont{Kociak}}, \bibnamefont{and}
  \bibinfo{author}{\bibfnamefont{F.~J.~G.} \bibnamefont{de~Abajo}},
  \bibinfo{journal}{Nature materials} \textbf{\bibinfo{volume}{18}},
  \bibinfo{pages}{1158} (\bibinfo{year}{2019}).

\bibitem[{\citenamefont{Pierce}(1947)}]{TWT}
\bibinfo{author}{\bibfnamefont{J.}~\bibnamefont{Pierce}},
  \bibinfo{journal}{Proceedings of the IRE} \textbf{\bibinfo{volume}{35}},
  \bibinfo{pages}{111} (\bibinfo{year}{1947}).

\bibitem[{\citenamefont{Martins and Silva}(2014)}]{eaccel}
\bibinfo{author}{\bibfnamefont{M.}~\bibnamefont{Martins}} \bibnamefont{and}
  \bibinfo{author}{\bibfnamefont{T.}~\bibnamefont{Silva}},
  \bibinfo{journal}{Radiation Physics and Chemistry}
  \textbf{\bibinfo{volume}{95}}, \bibinfo{pages}{78} (\bibinfo{year}{2014}).

\bibitem[{\citenamefont{Liu et~al.}(2017)\citenamefont{Liu, Xiao, Ye, Wang,
  Cui, Feng, Zhang, and Huang}}]{rf43a}
\bibinfo{author}{\bibfnamefont{F.}~\bibnamefont{Liu}},
  \bibinfo{author}{\bibfnamefont{L.}~\bibnamefont{Xiao}},
  \bibinfo{author}{\bibfnamefont{Y.}~\bibnamefont{Ye}},
  \bibinfo{author}{\bibfnamefont{M.}~\bibnamefont{Wang}},
  \bibinfo{author}{\bibfnamefont{K.}~\bibnamefont{Cui}},
  \bibinfo{author}{\bibfnamefont{X.}~\bibnamefont{Feng}},
  \bibinfo{author}{\bibfnamefont{W.}~\bibnamefont{Zhang}}, \bibnamefont{and}
  \bibinfo{author}{\bibfnamefont{Y.}~\bibnamefont{Huang}},
  \bibinfo{journal}{Nature Photonics} \textbf{\bibinfo{volume}{11}},
  \bibinfo{pages}{289} (\bibinfo{year}{2017}).

\bibitem[{\citenamefont{Sapra et~al.}(2020)\citenamefont{Sapra, Yang,
  Vercruysse, Leedle, Black, England, Su, Trivedi, Miao, Solgaard
  et~al.}}]{sapra2020chip}
\bibinfo{author}{\bibfnamefont{N.~V.} \bibnamefont{Sapra}},
  \bibinfo{author}{\bibfnamefont{K.~Y.} \bibnamefont{Yang}},
  \bibinfo{author}{\bibfnamefont{D.}~\bibnamefont{Vercruysse}},
  \bibinfo{author}{\bibfnamefont{K.~J.} \bibnamefont{Leedle}},
  \bibinfo{author}{\bibfnamefont{D.~S.} \bibnamefont{Black}},
  \bibinfo{author}{\bibfnamefont{R.~J.} \bibnamefont{England}},
  \bibinfo{author}{\bibfnamefont{L.}~\bibnamefont{Su}},
  \bibinfo{author}{\bibfnamefont{R.}~\bibnamefont{Trivedi}},
  \bibinfo{author}{\bibfnamefont{Y.}~\bibnamefont{Miao}},
  \bibinfo{author}{\bibfnamefont{O.}~\bibnamefont{Solgaard}},
  \bibnamefont{et~al.}, \bibinfo{journal}{Science}
  \textbf{\bibinfo{volume}{367}}, \bibinfo{pages}{79} (\bibinfo{year}{2020}).

\bibitem[{sm()}]{sm}
\emph{\bibinfo{title}{\uppercase{S}ee supplemental material for detailed proof
  of an electron beam's dynamic conductivity model as well as numerical
  simulation methods, along with influence of the outer radius of the
  cylindrical waveguide}}.

\bibitem[{\citenamefont{Cherenkov}(1934)}]{rf43b}
\bibinfo{author}{\bibfnamefont{P.~A.} \bibnamefont{Cherenkov}},
  \bibinfo{journal}{Dokl. Akad. Nauk} \textbf{\bibinfo{volume}{SSSR 2}},
  \bibinfo{pages}{451} (\bibinfo{year}{1934}).

\bibitem[{\citenamefont{Smith and Purcell}(1953)}]{smith1953visible}
\bibinfo{author}{\bibfnamefont{S.~J.} \bibnamefont{Smith}} \bibnamefont{and}
  \bibinfo{author}{\bibfnamefont{E.}~\bibnamefont{Purcell}},
  \bibinfo{journal}{Physical Review} \textbf{\bibinfo{volume}{92}},
  \bibinfo{pages}{1069} (\bibinfo{year}{1953}).

\bibitem[{\citenamefont{Jackson}(1999)}]{jackson1999classical}
\bibinfo{author}{\bibfnamefont{J.~D.} \bibnamefont{Jackson}},
  \emph{\bibinfo{title}{Classical electrodynamics}} (\bibinfo{year}{1999}).

\bibitem[{\citenamefont{Sydoruk et~al.}(2010)\citenamefont{Sydoruk, Kalinin,
  and Solymar}}]{sydoruk2010terahertz}
\bibinfo{author}{\bibfnamefont{O.}~\bibnamefont{Sydoruk}},
  \bibinfo{author}{\bibfnamefont{V.}~\bibnamefont{Kalinin}}, \bibnamefont{and}
  \bibinfo{author}{\bibfnamefont{L.}~\bibnamefont{Solymar}},
  \bibinfo{journal}{Applied Physics Letters} \textbf{\bibinfo{volume}{97}},
  \bibinfo{pages}{062107} (\bibinfo{year}{2010}).

\bibitem[{rf4()}]{rf45}
\emph{\bibinfo{title}{\uppercase{COMSOL}
  \uppercase{M}ultiphysics\textsuperscript{\textregistered} v. 5.4.
  www.comsol.com. \uppercase{COMSOL} \uppercase{AB}, \uppercase{S}tockholm,
  \uppercase{S}weden}}.

\bibitem[{egu()}]{egun}
\emph{\bibinfo{title}{\uppercase{STAIB} \uppercase{I}nstruments
  \uppercase{I}nc, www.staibinstruments.com. \uppercase{STAIB}
  \uppercase{I}nstruments, \uppercase{V}irginia, \uppercase{U}nited
  \uppercase{S}tates}}.

\end{thebibliography}


\begin{thebibliography}{11}
\bibitem{rf46s} U.  S.  Inan  and  M.  Golkowski,  Derivation  of  the  second moment of the Boltzmann equation, in Principles of Plasma Physics for Engineers and Scientists(Cambridge University Press, 2010) p. 261–262.
\bibitem{boltzmn2s}   O. Sydoruk, E. Shamonina, V. Kalinin, and L. Solymar, Terahertz instability of surface optical-phonon polaritons that interact with surface plasmon polaritons in the presence  of  electron  drift,  Physics  of  Plasmas 17,  102103(2010).
 \bibitem{boltzmn3s} S.   Riyopoulos,   Excitation   of   coupled   ion   lattice-streaming carrier modes in high mobility semiconductors, Physics of Plasmas 16, 033103 (2009).
 \end{thebibliography}
